\documentclass[twocolumns]{emulateapj}

\usepackage{graphicx}
\usepackage{chemfig}
\usepackage{tikz}
\usetikzlibrary{matrix,shapes,arrows,positioning,chains,fit}

\usepackage{amsmath,amssymb,amstext}
\usepackage[breaklinks,colorlinks,citecolor=blue,linkcolor=magenta]{hyperref}
\usepackage[all]{hypcap}
\usepackage{aas_macros}
\usepackage{natbib}
\usepackage{xcolor}
\usepackage{makecell}
\usepackage{adjustbox}

\newcommand{\ppm}{\ensuremath{ \vec{p}_m  }}
\newcommand{\fe}{\ensuremath{\mathrm{[Fe/H]}}}
\newcommand{\dfe}{\ensuremath{\nabla\mathrm{[Fe/H]}}}	
\newcommand{\Ro}{\ensuremath{R_0}}
\newcommand{\tRo}{\ensuremath{\tilde{\mathrm{R}}_0}}

\newcommand{\arexp}{\ensuremath{\alpha_{\mathrm{R_{exp}}}}}
\newcommand{\tsfr}{\ensuremath{\tau_{\tiny{SFR}}}}
\newcommand{\srm}{\ensuremath{\sigma_\mathrm{RM8}}}
\newcommand{\rfeh}{\ensuremath{R_{\fe =0}^{\mathrm{now}}}}
\newcommand{\tfeh}{\ensuremath{\tau_\mathrm{[Fe/H]}}}
\newcommand{\gfe}{\ensuremath{{\gamma_\mathrm{[Fe/H]}}}}
\newcommand{\mfe}{\ensuremath{\overline{\fe}}}			
\newcommand{\sfe}{\ensuremath{\mathrm{std(\fe)}}}		
\newcommand{\rold}{\ensuremath{R_\mathrm{old}}}			

\newcommand{\data}{\ensuremath{\{\fe , \tau , R\}}}

\newcommand{\fetau}{\ensuremath{\{\fe , \tau \}}}
\newcommand{\fetaui}{\ensuremath{\{\fe , \tau \}_i}}
\newcommand{\Ri}{\ensuremath{\{R\}_i}}

\newcommand{\pt}{\ensuremath{p(\tau ~|~\ppm )}}
\newcommand{\pRogt}{\ensuremath{p(\Ro ~|~\tau, \ppm )}}
\newcommand{\tpRogt}{\ensuremath{p(\tRo ~|~\tau, \ppm )}}

\newcommand{\pRgRot}{\ensuremath{p(R~|~\Ro , \tau, \ppm) }}
\newcommand{\tpRgRot}{\ensuremath{p(R~|~\tRo , \tau, \ppm) }}
\newcommand{\pfetgr}{\ensuremath{p(\fe , \tau~|~ R, \ppm )}}

\newcommand{\plik}{\ensuremath{p_{\mathcal{L}}\bigl ( \fetau ~|~ \{R\}, \ppm\bigr )}}
\newcommand{\pliki}{\ensuremath{p_{\mathcal{L}} \bigl ( \fetaui ~|~ \Ri, ~\ppm\bigr )}}

\newcommand{\posterior}{\ensuremath{p_{po}}}

\newcommand{\rom}{radial orbit migration}
\newcommand{\srom}{radius migration}

\DeclareMathOperator{\erf}{erf}

\shorttitle{Measuring Galactic Orbit Migration Efficiency}
\shortauthors{Frankel et al.}

\begin{document}\frenchspacing

\title{Measuring Radial Orbit Migration in The Galactic Disk}
\author{Neige Frankel\altaffilmark{1}, Hans-Walter Rix\altaffilmark{1}, Yuan-Sen Ting\altaffilmark{2,3,4}, Melissa~Ness\altaffilmark{5,6}, and David~W.~Hogg\altaffilmark{1,6,7,8}}
\altaffiltext{1}{Max Planck Institute for Astronomy, K\"onigstuhl 17, D-69117 Heidelberg, Germany}
\altaffiltext{2}{Institute for Advanced Study, Princeton, NJ 08540, USA}
\altaffiltext{3}{Department of Astrophysical Sciences, Princeton University, Princeton, NJ 08544, USA}
\altaffiltext{4}{Observatories of the Carnegie Institution of Washington, 813 Santa Barbara Street, Pasadena, CA 91101, USA}
\altaffiltext{5}{Department of Astronomy, Columbia University, 550 W 120th St, New York, NY 10027, USA}
\altaffiltext{6}{Flatiron Institute, 162 5th Avenue, New York, NY 10010, USA}
\altaffiltext{7}{Center for Cosmology and Particle Physics, Department of Physics, New York University, 726 Broadway, New York, NY 10003, USA}
\altaffiltext{8}{Center for Data Science, New York University, 60 5th Avenue, New York, NY 10011, USA}

%
%
%
%
%
%
\begin{abstract}
We develop and apply a model to quantify the global efficiency of \rom~among stars in the Milky Way disk. This model parameterizes the possible
star formation and enrichment histories, radial birth profiles, and combines them with a migration model that relates present-day orbital radii to birth radii through a Gaussian probability, broadening with age $\tau$ as $\srm~\sqrt{\tau/8{\mathrm{Gyr}}}$. Guided by observations, we assume that stars are born with an initially tight age--metallicity relation at given radius, which becomes subsequently scrambled by \rom, thereby providing a direct observational constraint on \rom~strength \srm. We fit this model with MCMC to the observed age--metallicity distribution of low-$\alpha$ red clump stars with Galactocentric radii between 5 and 14 kpc from APOGEE DR12, sidestepping the complex spatial selection function and accounting for the considerable age uncertainties. This simple model reproduces well the observed data, and we find a global (in radius and time) \rom~efficiency in the Milky Way of $\srm=3.6\pm 0.1$~kpc when marginalizing over all other model aspects. This shows that \rom~in the Milky Way's main disk is indeed rather strong, in line with theoretical expectations: stars migrate by about a half-mass radius over the age of the disk. The model finds the Sun's birth radius at $\sim 5.2$~kpc. If such strong \rom~ is typical, this mechanism plays indeed an important role in setting the structural regularity of disk galaxies.

\end{abstract}

\keywords{Galaxy: abundances --- Galaxy: disk --- Galaxy: evolution --- Galaxy: formation --- ISM: abundances --- stars: abundances}

%
%
%
%
%
%

\section{Introduction}
\label{sec:introduction}
To understand how disk galaxies formed and evolved \citep[e.g.,][]{MovdBWhite2010, schonrich_binney_2009b}, we need to understand how our Milky Way, typical disk galaxy, formed and evolved. In particular, we need to identify and characterize the processes setting the radial and vertical structures of the Galactic stellar disk in terms of stellar ages and abundances.

The present-day structure must at some level reflect both the global initial conditions such as the cold gas' total angular momentum and distribution, and the hierarchical merging during the early turbulent phases of the Milky Way's formation \citep{Bird2013,Stinson2013}. The stars' age distribution obviously reflects the overall star formation history of the Galaxy. In addition, the stars' photospheric element abundances trace the gradual enrichment of the Milky Way, which proceeded differently in different parts of the galaxy \citep[e.g.,][]{schonrich_binney_2009b}.

But for the last $\sim 8$~Gyr, the Milky Way's dynamical history has been quite quiescent, with the large majority of stars formed since then residing in a thin disk \citep[e.g.,][]{rix_bovy_2013,bland-hawthorn_gerhard_2016}. 
However, even in this quiescent regime, we cannot expect the stars' present-day orbits to reflect their birth orbits, as first detailed by 
\cite{sellwood_binney_2002} (SB02): there may be a great deal of dynamical evolution
on timescales longer than a dynamical time because the Galaxy is not axisymmetric, called ``secular evolution''.

In particular, \rom~has been recognized as a potentially very important process in both analytic and simulation work \citep{sellwood_binney_2002,Roskar2008,MinchevFamey2010}. Even if a star was born on a circular orbit, its present-day radius may differ from its birth radius for basically two reasons: first, a variety of perturbations in the in-plane or vertical direction may cause increasing epicycles, a process dubbed ``blurring'' by SB02 to refer to orbital heating. We know from the velocity dispersion in the Galactic disk that for ``middle-aged'' stars ($\sim 5$~Gyr), this leads to radial excursions of about 1~kpc. But SB02 emphasized another process, which
they dubbed ``churning'', that occurs in the presence of changing, fleeting or complex non-axisymmetric patterns (over-densities) such as spiral arms; these exert torques on stars, and lead to an effective change in a star's angular momentum or mean orbital radius, without inducing much ``blurring''. Here, we focus on the changes in the (instantaneous) orbital radius, and refer to this combined effect of ``churning'' and ``blurring'' as ``\rom'' or ``\srom'' throughout the present analysis.

N-body and cosmological simulations imply that \rom~is very important: the angular momentum (and hence orbit-size) may change of order unity for any
star over time-scales as short as a few Gyr \citep{kubryk_etal_2013,grand_etal_2016}. But to predict the actual 
degree of \srom~ in any galaxy {\sl quantitatively},
one would need to have an inventory of all the past spiral and bar 
pattern speeds and strengths.

There is well established observational evidence for the relevance of this process. In external galaxies, it makes predictions for the outermost radial density and age profiles of the stellar disk, which are in qualitative agreement with observations \citep[e.g.,][]{herpich_etal_2017,ruiz-lara_etal_2017}. In our Galactic disk, there is the remarkable, longstanding observation that there is no distinct age--metallicity relation of stars in the Solar neighbourhood (few 100~pc around the Sun); and that there is a wide spread of metallicities at the Solar radius \citep{edvardsson_etal_1993,casagrande_etal_2011}.
Both observations would be puzzling if stars -- at a given time and Galactocentric radius -- were born with a very small spread in metallicities. This is expected from both chemical evolution models \citep{Matteucci1989} and from observations of the interstellar gas and young stars in galaxies \citep[e.g.,][]{Przybilla2008}.

But if there is an important radial gradient in the metallicities (as
observed in the Milky Way, \citep[e.g.,][]{genovali_etal_2014}) then extensive \srom,
scrambling the orbital radii of stars while keeping their [Fe/H] unchanged, could explain the lack of an age--metallicity--(present-day) radius relation at given radius. This has been advocated and worked out by \cite{schonrich_binney_2009a,schonrich_binney_2009b,  roskar_etal_2008b,minchev_etal_2013}.  
They laid out a picture where 
three basic ingredients can explain the present-day orbit--age--abundance distribution of Galactic disk stars:
1) disk stars at all epochs and Galactocentric orbit radii were born with a well-determined metallicity (\fe $(\Ro,\tau)$), 2) there has always been an evolving outward metallicity gradient, 3) extensive subsequent mixing of orbital radii occurred.

Much of the best observational constraints on such \rom~in the Galaxy
stems from very local samples \citep{nordstrom_etal_2004,sanders_binney_2015}, with stars that have both abundance and age estimates; and indeed these analysis imply very effective \srom.
But if \rom~is a global phenomenon across the Galactic disk, then it calls for a ``global'' test, i.e. a test with observational data that cover Galactocentric radii that encompass a good fraction of the Galactic disk.

Here we propose {\sl measurement} of the average, or global, efficiency of \rom, based on data over a very wide Galactocentric radial range ($5\le R \le 14~\mathrm{kpc}$), with age estimates from spectroscopy. APOGEE \citep{majewski_etal_2017} spectra provide the first large  ($\sim 20,000$) sample with consistent age estimates, $\tau$ \citep{ness_etal_2016} across a large radial range in the Galaxy.
Qualitatively, the young stars $\tau\le 1$~Gyr show a well defined radial metallicity gradient (\fe~ decreasing outward), with a modest scatter in \fe  ~ at any given radius (see Figure \ref{fig_data_metallicity_radius_plane}). ``Old" stars ($\ge 10$~Gyr) show no discernable
metallicity -- radius relation, or at least enormous scatter in $p(\fe ~|~ R )$. The basic idea \citep{schonrich_binney_2009a,sanders_binney_2015} is that extensive \srom~ has largely erased the original radius--\fe-- age relation.

This approach is related to, but not the same as ``chemical tagging" \citep{freeman_bland-hawthorn_2002,ting_etal_2015a}, which aims at identifying stars that were born in the same cluster by their near-identical, detailed abundance patterns, even if they are now on widely different orbits. While stars from the same cluster were manifestly born at the same epoch and the same Galactocentric radius, the approach in the present analysis makes a different assumption: that stars born at the same epoch at the same Galactocentric radius have very similar [Fe/H] \citep[e.g.,][]{Przybilla2008}.

Any \srom~ over the course of a star's life is best thought of as a combination of diffusion of orbital angular momentum, or guiding radius (churning) and orbital heating (blurring), presuming it was born on a near-circular orbit. These are two distinct processes of different amplitudes, which can be measured separately using stellar angular momenta and radial action. But here, we focus on the stars' Galactocentric radii $R$ as a proxy, as these quantities are currently available with great fidelity and across a wide range of radii. We also restrict our analysis to stars with ages $\tau\le 8$~Gyr, as a model of gradual, secular orbit evolution may not be applicable to the earliest phases of the Galactic (thick) disk formation.

Here we construct a forward-model that incorporates the main processes that set the age- and abundance-dependent structure of the Galactic disk: the global star-formation history, inside-out growth, gradual chemical enrichment and \rom. In important aspects, this model draws on the ideas laid out in \cite{sanders_binney_2015}. We then compare this model to  APOGEE data, thereby constraining the strength of \rom~from data across the Galaxy. The data are presented in Section \ref{section_data}. The methodology is laid out in Section \ref{section_methodology}. We then present our results in Section \ref{section_results}  that quantitatively
constrain \srom~ and affirm how effective it seems to be in the Galaxy.  We conclude and comment in Section \ref{section_ccl}

%
%
%
%
%
%

\section{Data: APOGEE red clump giants}
\label{section_data}

\begin{figure*}
\includegraphics[scale=0.9]{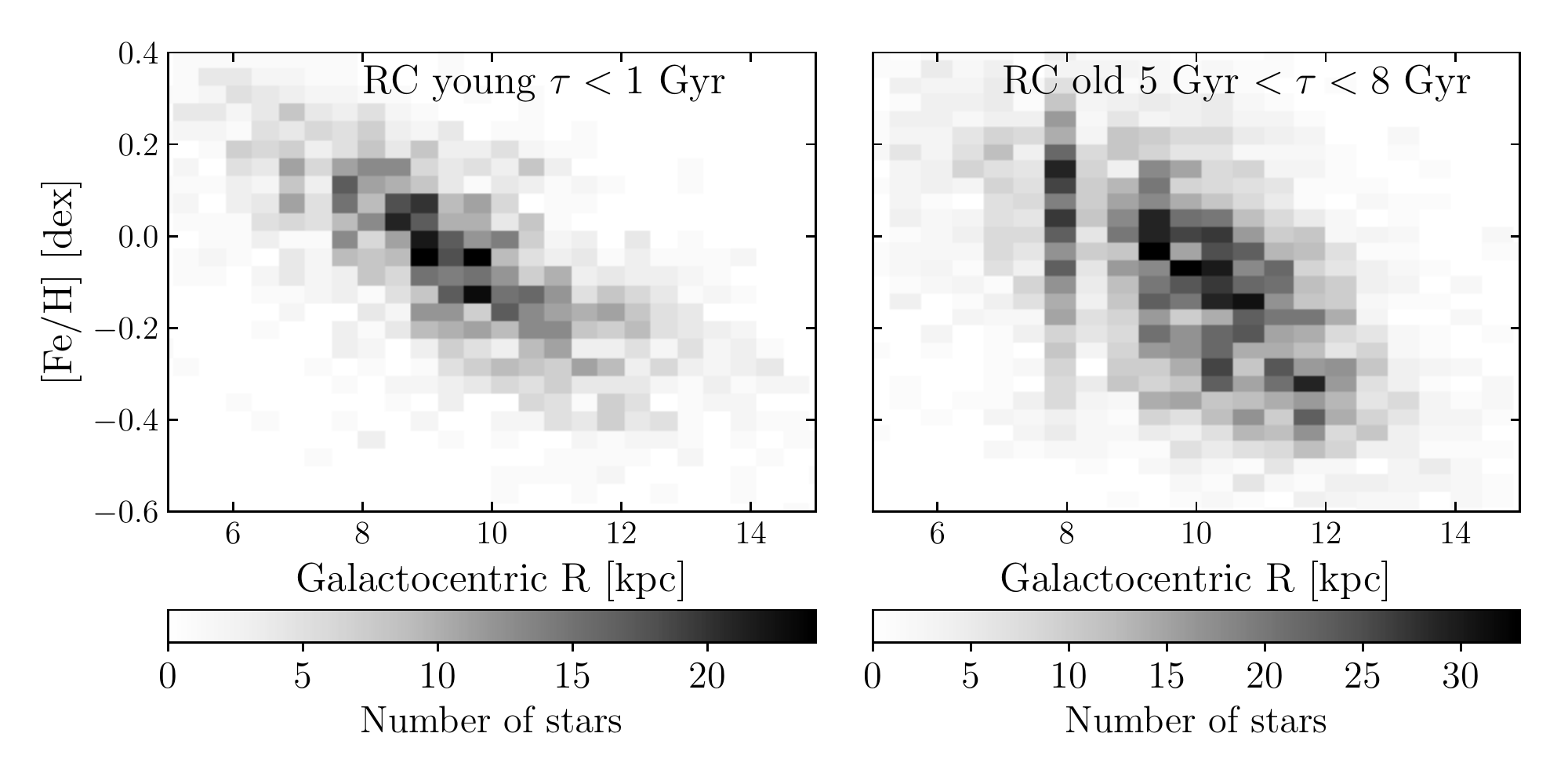}
\caption{\label{fig_data_metallicity_radius_plane} Number density distribution of RC stars in the plane of metallicity ([Fe/H]) and Galactocentric radius, for two ages bins: young stars (less than 1 Gyr, left), and older stars (measured age between 5 and 8 Gyr, right). Measurement uncertainties are of about 5\% in radius and 0.1 dex in metallicity. This Figure, adapted from \cite{ness_etal_2016}, shows that (1)
there is a well-defined metallicity gradient among young stars; (2) at a given metallicity, the (horizontal) spread in Galactocentric radii is larger for old stars than that for young stars (which we interpret and model as a consequence of \srom); and (3) the stellar density at different radii is dominated by the complex spatial selection function of APOGEE (e.g., the manifest over-density at 8 kpc, reflecting the location of the Sun).}
\end{figure*}

Global constraints on \srom ~of stellar orbits call for a sample of stars that covers a wide range in Galactocentric radii at low latitudes and with precise distances, and that has consistent [Fe/H] and age estimates.
The APOGEE \citep[Apache Point Observatory Galactic Evolution Experiment][]{majewski_etal_2017} sample of red clump giants
\citep{Alam_2015_apogeedr12,bovy_etal_2014} is, by design, very well suited for this purpose. 
Observing at near-infrared wavelengths for which dust is nearly transparent, the APOGEE spectrograph delivered spectra for giant stars with Galactocentric radii from $\sim$ 5 kpc to $\sim$ 14 kpc, as illustrated in Figure \ref{fig_data_metallicity_radius_plane}. Stellar parameters and abundances for this sample \citep[originally from APOGEE DR12,][]{Alam_2015_apogeedr12} were re-derived using \textit{The Cannon} \cite{ness_etal_2015}.
Importantly, consistent ages were derived by 
\cite{ness_etal_2016}, using the same data-driven approach to
calibrate spectroscopic age estimates to asteroseismic data; the spectroscopic age signature of 
red clump giants resides in the C and N abundances (at given [Fe/H]), reflecting mass (and hence age) dependent dredge-up \citep{Martig2016}. 
Uncertainties in metallicity are of about 0.05-0.10 dex, and those in ages ($\log{\tau}$) are 0.2~dex. 

Red clump giants are reliable standard candles, see for example \cite{girardi_2016}, with photometric distances precise to within 5\%. \cite{bovy_etal_2014} identified $\sim$ 20,000  red clump giants in APOGEE with a  contamination fraction between $\sim$ 3\% and 10\% by red giant branch stars. 


\begin{figure}
\includegraphics[scale=0.55]{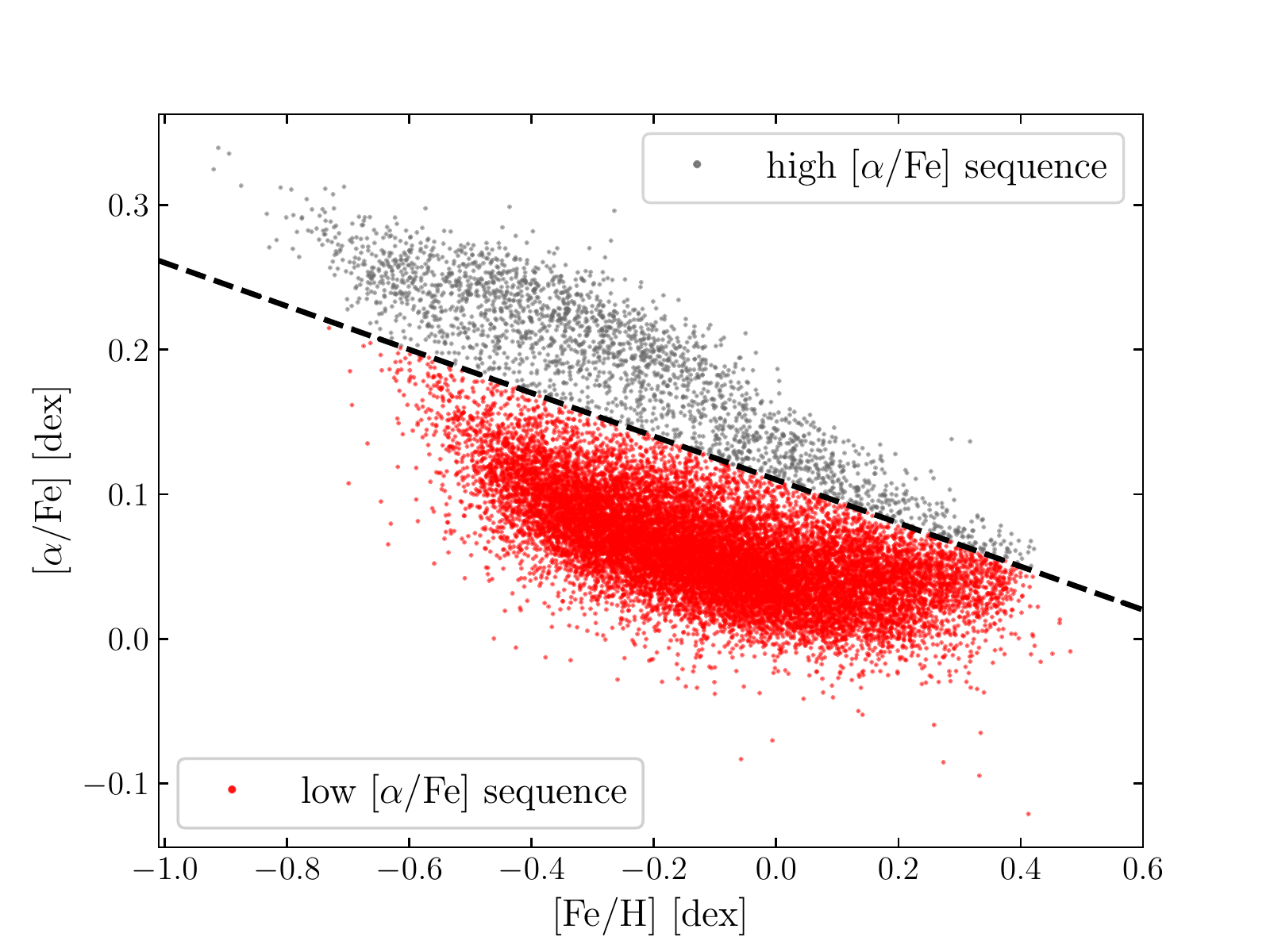}
\caption{\label{fig_low_alpha} Illustrating the abundance-based selection of
about 17,500 low-$\mathrm{[\alpha / Fe]}$ red clump stars among the 20,000 APOGEE red clump giants (in red, below the dashed line) for this study.
We focus on those ``thin disk" stars, as describing orbit evolution via gradual, secular \rom~may not be applicable to the turbulent early phases of Milky Way formation, when most high-$\mathrm{[\alpha / Fe]}$ presumably formed.}
\end{figure}

The above elements provide us with a set of about 20 000 data \data , and their uncertainties. For our modelling at hand, it seems sensible to apply a few more cuts to the sample. As we are interested in \rom~as the possibly dominant orbit evolution process in the more quiescent phase of the Galactic disk evolution (the last $\sim$8~Gyr), we eliminate stars with high  $[\alpha / \mathrm{Fe}]$, as illustrated by the grey dots in Fig \ref{fig_low_alpha}. Additionally, we select stars well in the Galactic plane with altitude $|z| < 1$ kpc.

\subsection{Sidestepping the complex spatial selection function}
Given a set of data \data , the obvious approach would be to construct a parameterized
model to predict $p(\data |~ \ppm)$, where $\ppm$ are various model parameters describing the possible evolution histories of the Galactic disk  (see Section~\ref{section_methodology}) including \rom. 
But such direct comparison of model predictions to data requires to account for the selection function: the probability that any star in the sky enters the survey catalog, given its physical properties. 
In the case of the APOGEE data at hand, the selection function is (inevitably) complex: stars must (1) belong to the red clump population, and (2) be the pointing directions of APOGEE and have color and magnitudes to fit the APOGEE survey selection. 

Firstly, the number of red clump stars per unit mass of a stellar population is a strong function of age \citep{Girardi2001}. \cite{bovy_etal_2014} have calculated with stellar evolutionary models the relative fraction of stars which are in this evolutionary stage in function of their age for a flat star formation history (this is illustrated by the dashed line in Figure \ref{fig_age_distribution}).

Secondly, the APOGEE spatial selection function was shown to be a complex function \citep{bovy_rix_etal_2016}. The consequences of spatial distribution on the radial distribution of the APOGEE red clump sample used in our study is visible in Figure \ref{fig_data_metallicity_radius_plane} where there is, for example, an over-density of stars observed at the position of the Sun ($\approx$ 8 kpc). Therefore, we opt not to model this complex distribution. Instead, we work only with the age--metallicity distribution given stellar radii $p(\{\fe , \tau \} ~|~\{R \})$. The advantage is that the model construction is technically simpler and more robust; but not all of the information contained in the data is used. In particular, we are not exploiting the present-day radial distribution of stars in the Milky Way disk $p(\{R \})$.

%
%
%
%
%
%
\section{A model for the Galactic disk evolution, including radial orbit migration}
\label{section_methodology}

We now lay out a simple parameterized model for the age--abundance--radius structure of the Galactic disk of low-[$\alpha$/Fe] stars.
This model specifies different formation and evolution aspects: when, and at what metallicity stars were born, with which radial profile they were born, and how much they migrated, ultimately
predicting the joint distribution $p(\fe,\tau, R)$ and $p(\{\fe , \tau \} ~|~\{R \})$. In many ways, this model draws on the approach laid out by \cite{sanders_binney_2015}. 

We start by stating the main assumptions underlying our model. We then specify the individual model aspects, each described by a set of functional forms, which result in a vector of model parameters, $\ppm$. We then combine these aspects to predict the age-metallicity distribution at any given
Galactocentric radius, $p(\fe , \tau ~|~R, \ppm)$, which allows us to calculate the data likelihood for the APOGEE sample given any $\ppm$ and apply Bayes' theorem to infer the posterior probability function for the model parameters, given the data
\begin{equation}\begin{split}
p_{po}(\ppm ~| ~\data &)  =  \plik  \\
& \times p_{pr}(\ppm ) / p_{pr}(\fetau ),
\end{split}
\label{eq_parameterposterior}
\end{equation}
with $p_{po}$ the posterior probability density function of the model parameters, $p_{pr}(\ppm )$ our prior knowledge on the model parameters, and $p_{pr}(\fetau )$ the evidence.
Such inference operation requires to account for data uncertainties. We assume in the present study that the uncertainties in $R$ and \fe ~are negligible (red clump stars have $\sim 5\%$ and $\sim 0.05-0.1$ dex uncertainties in distance and \fe~ respectively). We presume that the uncertainties in  $\log{\tau}$ dominate and are described by a Gaussian with 
$\sigma_{\log{\tau}}=0.2$~dex \citep{ness_etal_2016}.

\subsection{Basic model assumptions \label{section_assumptions}} 
In order to describe the evolution of the Galactic disk with a parametrized model, we made several assumptions on the nature and strength of the processes at play. Obviously, the astrophysical inferences from our modelling are only as valid as the assumptions.
\begin{itemize}
\item We assume that the metallicity \fe ~of the interstellar medium has negligible variations with azimuth; this is perhaps the strongest assumption involved in the modelling. This assumption is supported by observations of young stars in the Galaxy \citep[e.g.,][]{Luck2006,Przybilla2008,genovali_etal_2014}.
Azimuthal variations in rapidly produced $\alpha$-elements have been claimed
\citep{Ho2017}, but those in [Fe/H] should be less strong.
\item We do not treat or model explicitly the vertical structure of the Milky Way disk, though there are of course vertical (populations) gradients in it \citep[e.g.][]{ness_etal_2016}. 
\item Secular evolution has been the dominant orbit evolution effect for the past 8 Gyr, which implicitly assumes that the Milky Way had a relatively quiescent life for the past 8 Gyr. We therefore restrict our analysis to stars younger than 8 Gyr, neglecting possible recent external interactions that could be responsible for shaping the Milky Way disk.
\end{itemize}
It follows from these assumptions that we model \rom~as the only mechanism responsible for the scatter in age--metallicity at given radius. In this work, we interpret all scatter with \srom, and therefore provide an upper limit on its strength, which should reflect the distance over which stars have migrated radially.

\subsection{Functional Forms for the Different Aspects of the Model}
\label{subsec_model}

In the following, we use the assumptions stated above and lay out our adopted functional forms for different aspect of the Galactic disk's formation and evolution: the distributions of (1) the global disk star-formation rate, (2) birth radii distribution as a function of time, (3) birth metallicities at a given epoch and radius, and (4) the strength of \rom. We summarize these functional forms in Table \ref{table_model_description}. These functions are combined to produce Eq \ref{eq_parameterposterior}, from which we can sample the posterior probability distribution function of the parameters \ppm.

\subsubsection{Star formation history and the age distribution of red clump stars}

We parameterize the possible age distribution of red clump stars by
\begin{equation}
\label{eq_age_dist}
\pt \equiv c_1 \cdot \mathrm{SFH}(\tau, \ppm) \cdot f_{RC}(\tau),
\end{equation}
where SFH is the star formation history of the Milky Way thin disk, $f_{RC}$ is the relative mass of stars at the red clump stage, and the normalization requires
\begin{equation}
 c_1^{-1} \equiv \int_0^{\tau_m} \mathrm{SFH}(\tau, \ppm) f_{RC}(\tau)~d\tau .
\end{equation}

The star formation history (SFH) of the Milky Way thin disk is thought to be extended in time \citep{bland-hawthorn_gerhard_2016} and is manifestly still ongoing. This motivates our choice \citep{mo_etal_2010} to conventionally parametrize the star formation rate in the Milky Way disk as a slowly decreasing exponential with time, for which we fit the exponential decay time-scale \tsfr. This is a simplification of the \cite{sanders_binney_2015} model, who go further in detail and include thick disk star formation. We write the star formation history
\begin{equation}
\label{eq:p(tau|vecpar)}
\mathrm{SFH}(\tau, \ppm) = \exp{ \bigl [ -(\tau_m-\tau)/\tsfr } \bigr ],
\end{equation}
where $\tau_m$ is the maximum disk age, set to $12 $~Gyr; $\tsfr$ is the model parameter setting the star formation history, and is to be fit (i.e. it is an element of \ppm ).

\begin{figure}
\includegraphics[scale=0.55]{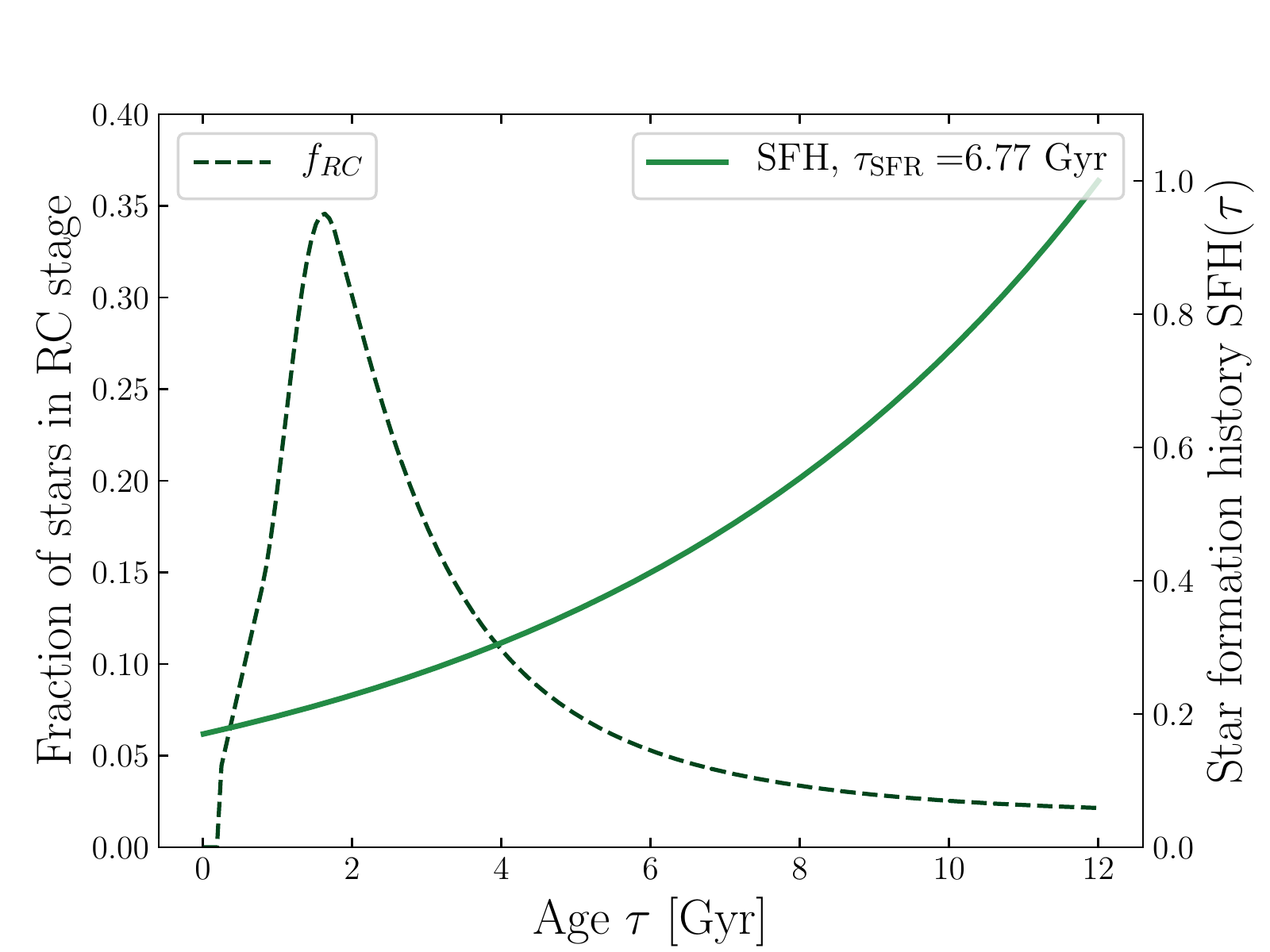}
\caption{\label{fig_age_distribution} Model for the global star formation history and the
age distribution of red clump stars. We assume that the global star formation history of the (low-$\alpha$) Galactic disk can be described (see Eq~\ref{eq:p(tau|vecpar)})
by a model family SFR $\propto \exp{(-t/\tsfr)}$,
illustrated by the solid line for a star formation time-scale $\tsfr = 6.8$ Gyr. 
The dashed line shows the theoretically expected relative number of red clump stars per unit mass for a constant star formation history. The normalized product of these two functions gives the current age distribution of red clump stars.}
\end{figure}

The expected number of red clump stars per unit stellar mass, $f_{RC}(\tau)$, is a distinct function of age (and a weaker function of metallicity); it has been derived and parametrized in eq (11) of \cite{bovy_etal_2014}. We illustrate $f_{RC}(\tau)$ in Figure \ref{fig_age_distribution} (dashed line) together with one particular choice of a star formation history SFH (solid line).

\subsubsection{Radial Birth profile and inside-out growth}
\begin{figure}
\includegraphics[scale=0.55]{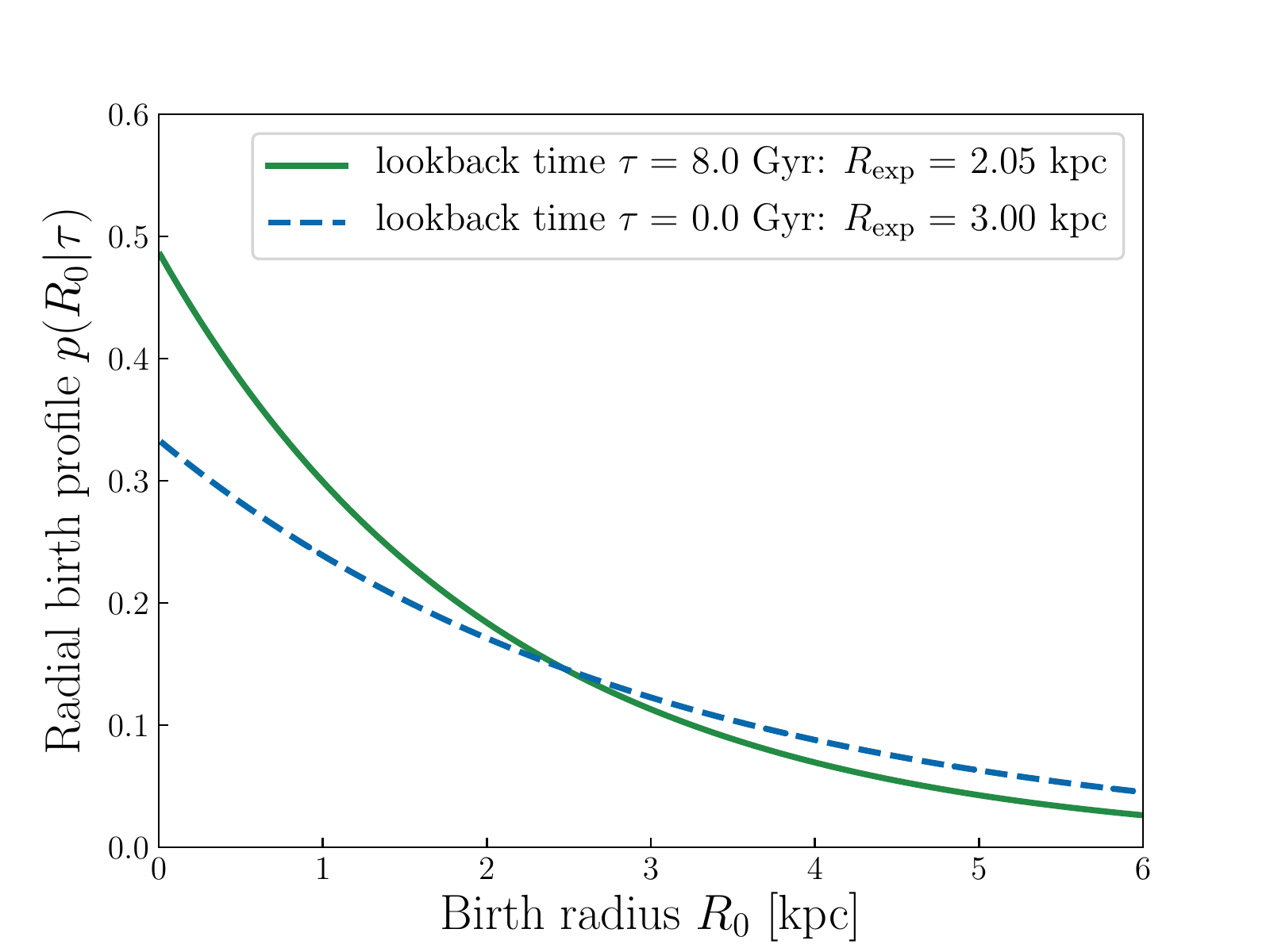}
\caption{\label{fig_radial_birth_profile} Illustration of the models for the birth-radius distribution of stars. At any given point in time, the radial birth profile is assumed exponential, with a scale-length growing with time to reflect inside-out growth (here, $\arexp = 0.3$, see Eq~\ref{eq:insideout}). Stars were born in more centrally concentrated regions 8 Gyr ago (solid line) with a radial scale length of 2 kpc, which is about 30\% smaller than today's assumed birth scale length of 3 kpc (dashed line).}
\end{figure}

We presume that disk stars are born on near-circular orbits near the mid-plane of the disk. The sizes of their orbits is determined by the angular momentum of the gas from which they formed. We therefore need to parametrize the radial profile of the star-forming gas in the Galactic disk at any time. The Galactic disk is thought to build from inside-out, as gas of first low then higher angular momentum cools and falls into the potential of the dark matter halo \citep{white_frenk_1991,mo_mao_white_1998, munos_mateos_etal_2007,fraternali_2012}. This inside-out growth is thought to play a determining role in the gas and stars metallicity profile \citep{schoenrich_mcmillan_2017}, so it is important to incorporate this aspect into our disk model. We parametrize the possible radial birth profile of stars at any given epoch as a decreasing exponential with Galactocentric radius, with a scale-length $R_\mathrm{exp}$,
\begin{equation} \label{eq:eq_radial_birth_profile}
\pRogt  = \exp{\bigl ( -\Ro / R_\mathrm{exp}(\tau)\bigr ) } / R_\mathrm{exp}(\tau).
\end{equation}
We then parameterize inside-out growth by allowing the scale-length to increase (linearly) with time,
\begin{equation}\label{eq:insideout}
R_\mathrm{exp}(\tau)=3~\mathrm{kpc}~\Bigl( 1-\arexp\bigl ( \frac{\tau}{{8 \mathrm{~Gyr}}}\bigr ) \Bigr ).
\end{equation}
The relative size of the disk today and at early times is set by the free parameter to be fit
$\arexp$ (Eq~\ref{eq:insideout}),
bound to the interval $[0,1]$ with the current star-forming disk scale-length set to $R_\mathrm{exp} (\tau = 0) = 3$ kpc. 
Note that we do not attempt to model the radial profile of the disk beyond 8 Gyr ago, because we deem our secular evolution model inapplicable at such early epochs.
The radial scale-length of the Milky Way stellar disk is not well constrained (see \cite{bland-hawthorn_gerhard_2016} for a review). It was shown that such scale-length varies with stellar populations \citep{bovy_etal_2012}. We adopt here the suggested value for the younger stars (in the chemical sense: with low $[\alpha /\mathrm{Fe}]$) in the disk of $\sim 3$ kpc from \cite{bovy_etal_2012}, to model the present-day star-forming gas profile. The possible distributions of stars at birth 8 Gyr ago and today are shown in Figure \ref{fig_radial_birth_profile} for a specific choice for \arexp.
\subsubsection{Metallicity--radius--age relation} \label{subsection_metallicity_radius_age}
\begin{figure}
\includegraphics[scale=0.55]{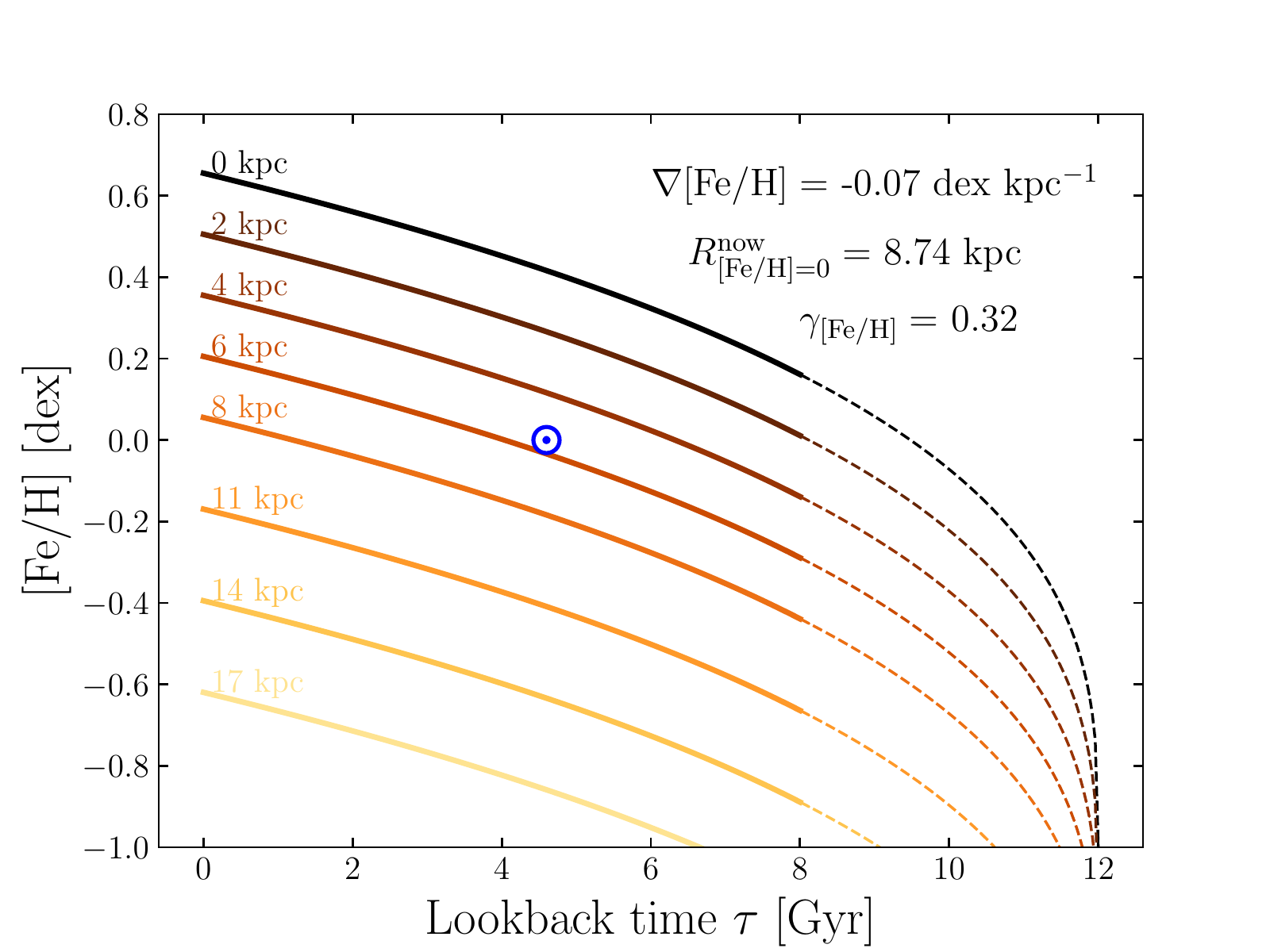}
\caption{\label{fig_age_metallicity} Model family for the ``chemical enrichment", i.e. the relation between age, birth radius and metallicity, shown for eight different birth radii and 
a fiducial set of model parameters (see inset and Eq~\ref{eq:eq_Fe_Ro_t}). At all times there is a 
radial metallicity gradient ([Fe/H] decreases towards larger radii), and at any radius (line of a given color) the interstellar medium gets enriched with time. The position of the Sun in this plane is indicated by the blue $\odot$ marker. Combinations of ages and metallicities above the black 0 kpc line would be deemed unphysical by the model.}
\end{figure}

We also need to specify with what [Fe/H] stars were born at time $\tau$ ago at Galactocentric radius $\Ro$.
At present, disk stars in the Milky Way are born with a tight relation between their birth radius and their metallicities. This is qualitatively seen in data: young sub-populations (e.g., Cepheids, \cite{genovali_etal_2014}) of a given \fe ~cover a small range of galactic radii. 
Open clusters metallicity spreads were shown to be about 0.03 dex \citep{bovy_2016,ness_etal_2017,ting_etal_2018}. This motivates our assumption that the metallicity profile of the interstellar medium (and hence the metallicity stars have at birth) can be modelled at any time through a tight relation. Following the general reasoning of \cite{sanders_binney_2015} who approximate the output of a simulation of \cite{schonrich_binney_2009a}, we describe the metallicity profile in the star-forming gas disk as the product of a radial profile, and of a term describing the time dependency of chemical enrichment
\begin{equation}\label{eq:eq_Fe_Ro_t}
\begin{split}
\fe = & F_m - (F_m + \dfe \rfeh) f(\tau )   \\ &+ \dfe  R.
\end{split}
\end{equation}
Here,
$F_m=-1$ dex is the minimum metallicity at the center of the disk, which we assume fixed, 
$\dfe$ is the interstellar medium metallicity gradient  in $\mathrm{dex}~ \mathrm{kpc}^{-1}$, is negative, and is to be fit with the other parameters in \ppm. It is presumed constant in radius and time, although the metallicity gradient may have evolved over the life time of the Galactic disk \citep{minchev_2018}. We discuss the possible impact of the assumed form for the metallicity profile in Section \ref{section_results}, where different expressions are tested. We expect young stars across the Galactic disk to provide the strongest constraints on this model parameter. Then, 
\rfeh ~is the radius at which the present-day (birth) metallicity is solar (\fe = 0). We expect this parameter to be constrained by the current radii of the youngest red clump stars of solar metallicity.
We assume the time dependency of the enrichment to follow the power law
\begin{equation*}
f(\tau ) = \left( 1 - \frac{\tau}{\tau_m} \right )^\gfe 
\end{equation*}
with the parameter (to fit) \gfe ~controlling the time dependency of chemical enrichment with time: linear if \gfe ~is 1, and faster at early times if  \gfe~is less.
Overall, this encapsulates that there is a metallicity gradient in the interstellar medium in the disk, and that enrichment proceeded gradually over time, as illustrated in Figure \ref{fig_age_metallicity}.

With this parametrization, we now assume that there is an exact birth metallicity at a given stellar age $\tau$ and birth radius $\Ro$, i.e.
$p \bigl (\fe~|~ \Ro , \tau , \ppm \bigr )$ is a $\delta$-function at the value of \fe~ that satisfies Eq \ref{eq:eq_Fe_Ro_t}.
To study \rom~($R - \Ro $), we use this functional form of the metallicity profile of the interstellar medium as a function of time to find stellar birth radii, given stellar metallicities and stellar ages. In other words, we invert the age--metallicity relation in Equation \ref{eq:eq_Fe_Ro_t} and construct the inverse relation $\tRo (\fe , \tau)$, which is a $\delta$-function in $\Ro$, centered on:
\begin{equation}\label{eq_Ro_ana}
\tRo = \frac{\fe - F_m + (F_m + \dfe \rfeh) f(\tau)}{\dfe}
\end{equation}
Such inversion requires
\begin{equation*}
\label{eq_R0_positive}
\tau \leq \tau _\mathrm{max}(\fe , \ppm),
\end{equation*}
for $\tRo$ to be positive. Here, $\tau _\mathrm{max}(\fe  , \ppm )$ is the maximum stellar age deemed physical by our model evaluated for \ppm, given a metallicity \fe. Solving the inequality $\tRo (\fe, \tau) > 0$ for $\tau$ at a given metallicity in Eq \ref{eq_Ro_ana},
\begin{equation}
\begin{split}
\tau _\mathrm{max} & (\fe, \ppm )  = \\ 
& \tau _m \Bigl( 1- \frac{\fe - F_m}{F_m - \dfe \rfeh}   \Bigr)^{1/\gfe}
\end{split}
\end{equation}
where we used the assumption that the metallicity gradient in the star-forming gas is always negative; [Fe/H] decreases outward. This inequality can be visualized in Figure \ref{fig_age_metallicity}: combinations of \fe ~and $\tau $ above the 0 kpc line are deemed unphysical. This condition, that is a function of \ppm, will therefore provide strong constraints on the parameters to fit in the age--metallicity -- birth radius relation, in particular on \gfe.

\subsubsection{Radial orbit migration}

\begin{figure*}
\centering
\includegraphics[scale=0.9]{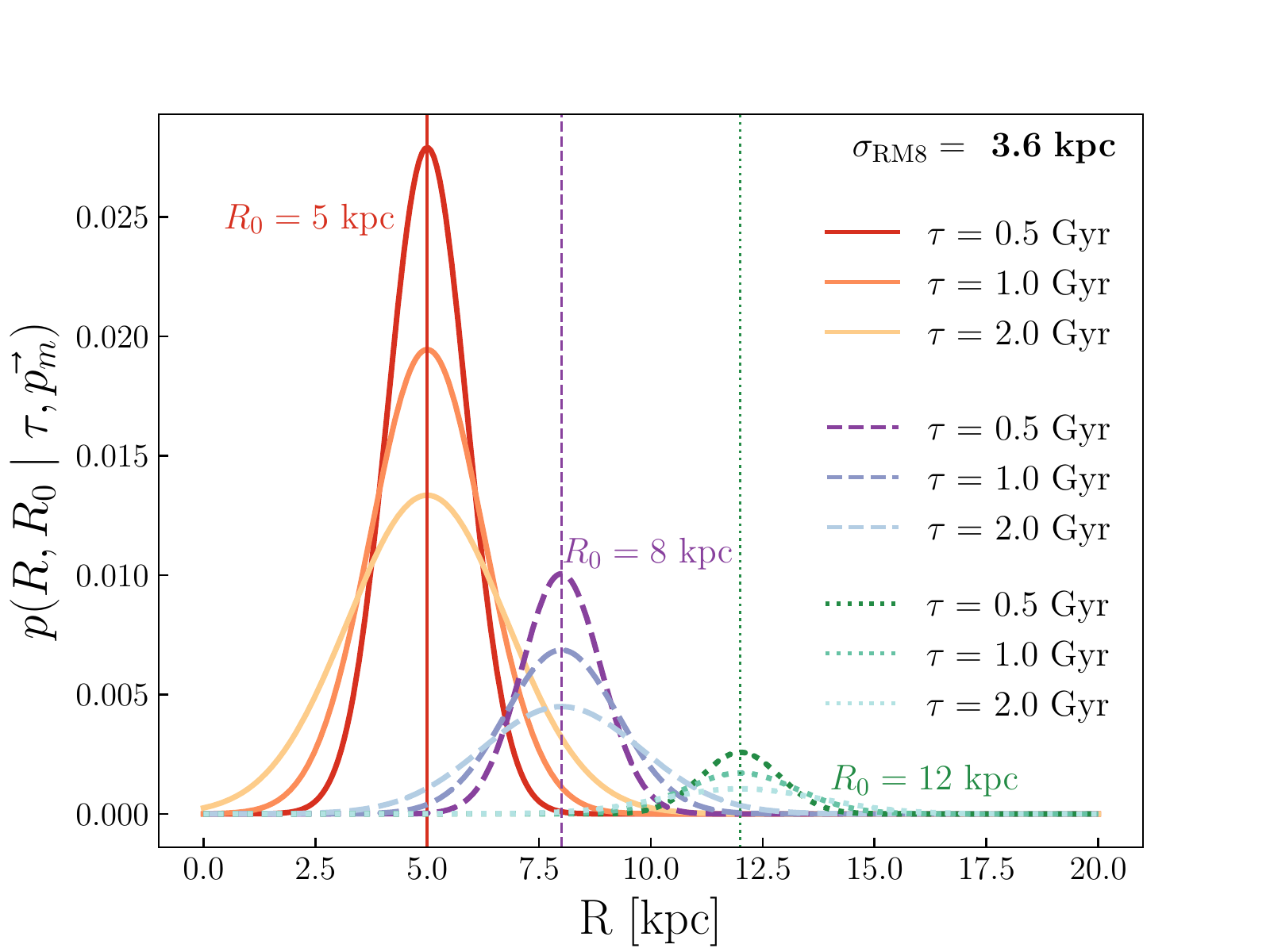}
\caption{\label{fig_radial_distributions} Simple model family for \rom,
illustrated by the orbit radius probability distributions of stars born respectively at 5, 8 and 12 kpc (plain red, dashed purple and dotted green) as a function of time after birth: 0.5, 1 and 2 Gyr (darker to lighter). The \rom~strength \srm~ to be fit (Eq~\ref{eq:eq_radial_migration}, here 3.6 kpc)
determines the rate at which the distributions broaden with time. Near the center of the Galactic disk, the distributions become asymmetric because stars have a null probability to migrate to negative radii (e.g., discontinuity of the yellow line at $\tau = 2$ Gyr and $R =0$ kpc). The distributions are modulated by the exponential radial birth profile, with an inside-out scale parameter \arexp ~ of 0.3, and the current disk scale-length 3kpc.}
\end{figure*}
%

\begin{table*}
\caption{\label{table_model_description}Summary of the important aspects of the model.}
\begin{tabular}{llccr}
\hline

\adjustbox{valign=t}{\makecell[l]{Question tackled \\by the model}}
& \adjustbox{valign=t}{\makecell[l]{Describing \\model parameter}}
&\adjustbox{valign=t}{\makecell[l]{Parameter \\ of \ppm to fit}}
&\adjustbox{valign=t}{\makecell{Relevant appearance\\ in the parametrized equations}} 
&\adjustbox{valign=t}{\makecell[r]{Model aspect \\ Eq reference}} \\
\hline
When did stars form? 
& Star formation timescale & 
\parbox{1cm}{
\begin{equation*}
\frac{\tsfr}{\mathrm{Gyr}}
\end{equation*}} & 
\parbox{6cm}{
\begin{equation*}\mathrm{SFH}(\tau, \ppm) = \exp{ \bigl [ -(\tau_m-\tau)/\tsfr } \bigr ] 
\end{equation*}} & 
\adjustbox{valign=t}{\makecell[r]{Star formation \\history\\Eq \ref{eq:p(tau|vecpar)}}} \\
Where did stars form? 
& \adjustbox{valign=t}{\makecell{Relative size of the disk \\ at birth and present-day}} & \arexp &
\parbox{5cm}{
\begin{equation*} \pRogt  \propto \exp{\bigl ( -\Ro / R_{exp}\bigr ) } 
\end{equation*}} & 
\adjustbox{valign=t}{\makecell[r]{Inside-out growth \\Eq \ref{eq:eq_radial_birth_profile}}}\\
& & & $R_{exp}\propto \bigl (1-\arexp\cdot  \frac{\tau}{8 \mathrm{Gyr}}\bigr )$ & \\
\hline
\adjustbox{valign=t}{\makecell[l]{With what \fe \\ were stars born? }}
&\adjustbox{valign=t}{\makecell[l]{ Present-day radius \\of solar metallicity \\in ISM }}&
\parbox{1cm}{
\begin{equation*}
 \frac{\rfeh}{\mathrm{kpc}}
 \end{equation*}}&
\parbox{6.5cm}{
\begin{equation*}
\begin{split}
&\fe =  F_m + \dfe  R \\ & - (F_m + \dfe \rfeh) f(\tau )  
\end{split}
\end{equation*}
}
& \adjustbox{valign=t}{\makecell[r]{Tight \\age-metallicity \\ relation at birth\\ Eq \ref{eq:eq_Fe_Ro_t}}} \\
& \adjustbox{valign=t}{\makecell[l]{Metallicity gradient\\ in the ISM}} &
\parbox{1cm}{
\begin{equation*}
\frac{\dfe}{\mathrm{dex.kpc^{-1}}}
\end{equation*}} & & \\
& \adjustbox{valign=t}{\makecell[l]{Enrichment \\power law index}} & $\gamma _\fe$&  
\parbox{5cm}{
\begin{equation*} 
f(\tau) =  \left( 1 - \frac{\tau}{\tau_m} \right )^{\gamma_\fe}
\end{equation*}}
& \adjustbox{valign=t}{\makecell{Chemical\\ enrichment}} \\
\hline
\adjustbox{valign=t}{\makecell[l]{How far did stars \\ orbit migrate \\over the disk life time?}}
& \adjustbox{valign=t}{\makecell[l]{Diffusion scale length \\ migration distance \\ over the past 8 Gyr}}&
\parbox{1cm}{
\begin{equation*} 
\frac{\srm }{\mathrm{kpc}} 
\end{equation*}}&
\parbox{5cm}{
\begin{equation*}
\pRgRot \propto \exp{\Bigl (-\frac{(R-\Ro)^2}{2~\srm^2~\tau/8 \mathrm{Gyr}}\Bigr )}
\end{equation*}
}
&\adjustbox{valign=t}{\makecell[r]{Diffusion in radius, \\ radial migration\\Eq \ref{eq:eq_radial_migration}}} \\
\hline
\end{tabular}
\end{table*}
%
%
We now introduce the central part of our model: \rom~in order to quantify how far stars move from their birth radii as a function of their age. Theoretical and observational arguments suggest that \rom~can be modelled as a diffusion process. \cite{sellwood_binney_2002} first demonstrated that non-axisymmetric structures such as spiral arms can, through repeated and transient torques on stars at co-rotating with them, induce large changes in their angular momenta. Further simulations confirmed this diffusion aspect of radial migration \citep{schonrich_binney_2009a, brunetti_etal_2011}. Qualitatively, data show that at a fixed metallicity, a spread in stellar radii increases with stellar ages. This is qualitatively evident in the different \fe--$R$ spread between the two panels in Figure \ref{fig_data_metallicity_radius_plane}. Motivated by these arguments, we follow \cite{sanders_binney_2015} and adapt their parametrization to Galactocentric radius coordinate.
In its simplest form, a solution to the diffusion equation in radius gives the following probability for a star to be currently at a Galactocentric radius $R$, given that it was born at \Ro ~a time $\tau $ ago:
\begin{equation} \label{eq:eq_radial_migration}
\pRgRot = c_3~ \exp{\Bigl (-\frac{(R-\Ro)^2}{2 ~\srm^2~\tau/{8 \mathrm{~Gyr}}}\Bigr )},
\end{equation}
where $\srm$, the \rom~strength (our main astrophysical goal, to fit), represents the extent of \rom~for a star after 8 Gyr (the width of the Gaussian function in Equation \ref{eq:eq_radial_migration} at age $ \tau = 8 $ Gyr). As its age increases, the probability for a star to be on a different orbit than its birth orbit increases, because it had more time to radial migrate. An illustration of the radial spread of different orbits with, for example, \srm ~ = 3.6 kpc is shown in Figure \ref{fig_radial_distributions}, where the distributions are modulated by the radial birth profile across the disk. 
Finally, the normalization constant $c_3$ satisfies
\begin{equation*}
c_3^{-1}=\srm \sqrt{\frac{\pi}{2} ~ ~\frac{\tau}{{8 \mathrm{~Gyr}}}}\cdot \biggl ( \erf \Bigl (\frac{\Ro}{\srm ~\sqrt{2}\sqrt{\tau/{8 \mathrm{~Gyr}}}}\Bigr ) +1 \biggr ),
\end{equation*}
to ensure that stars do not migrate to negative radii with the normalizing error function.

In this most restricted form, the only free parameter describing \rom~is $\srm$.

\subsection{Constructing the Data Likelihood Function}

We use the above elements to build a parameterized model that predicts the joint distribution \pfetgr ~at a given Galactocentric radius $ R $ for the low-$\alpha$ Galactic disk.
%
%
\begin{figure}
\centering
\begin{tikzpicture}
\tikzstyle{main}=[circle, minimum size = 12mm, line width=0.4mm, draw =black!80, node distance = 12mm]
\tikzstyle{err}=[circle, minimum size = 2mm, thick, draw =black!100, node distance = 12mm]
\tikzstyle{connect}=[-latex, thick]
\tikzstyle{box}=[rectangle, draw=black!100]
  \node[main, fill = white!100] (enrichment) [label=center:$\theta_\fe$] { };
   \node[main, fill = white!100] (alpha) [right=of enrichment, label=center:$\arexp$] { };
  \node[main, fill = white!100] (sigma) [right=of alpha, label=center:$\srm$] { };
 \node[main, fill = white!100] (R0) [below=8mm of alpha, label=center:$R_{0i}$] { };
\node[main, fill = white!100] (tau) [below=8mm of R0, label=center:$\tau '_i$] { };

  \node[main, fill = black!10, dashed] (R) [below=50 mm of sigma ,label=center:$R_i$] {}; 
  \node[main, fill = black!10] (feh) [below=50 mm of enrichment,label=center:$\fe _i$] { };
  \node[main, fill = black!10] (taui) [below=50mm of alpha,label=center:$\tau_i$] { };
  \node[err, fill = black!100](err)[below=of taui, label=right:$\sigma _\tau$]{};
  \node[main, fill=white!100](tsfr)[left=of err, label=center:$\tsfr$]{};
  \path (enrichment) edge [connect] (feh)
		(sigma) edge [connect] (R)
		(alpha) edge [connect] (R0)
		(tsfr) edge [connect] (tau)
		(tau) edge [connect] (R0)
		(tau) edge [connect] (R)
		(R0) edge [connect] (R)
		(R0) edge [connect] (feh)
		(tau) edge [connect] (feh)
		(tau) edge [connect] (taui)
		(err) edge [connect] (taui);
  \node[rectangle, thick, inner sep=0.mm, fit= (R) (taui),label=below right:{$ i=1,...,N $}, xshift=1.5mm] {};
  \node[rectangle, inner sep=5mm, draw=black!100, fit = (feh) (R) (taui) (R0)] {};
\end{tikzpicture}

\caption{\label{fig_graphical_model} Probabilistic graphical model for the joint distribution $p(\{\fe_i, \tau_i, R_i \} ~|~ \ppm)$ for the  $< 8$ Gyr Milky Way (thin) disk. Our likelihood is the ratio between this model and the model for  $p(\{ R_i \} ~|~ \ppm)$ presented in Figure \ref{fig_graphical_model_radius}. The observed quantities are in grey circles and model parameters are in white circles. The present-day Galactocentric radius $R_i$ is in a dashed circle as a reminder that the final likelihood does not predict the present-day observed radial distribution of red clump stars. The filled black dot represents a fixed quantity, here the assumed age errors, from \citep{ness_etal_2016}.The $\theta _\fe$ circle represents the three enrichment parameters {\rfeh, \gfe, \dfe}.~ \Ro ~are birth radii, $\tau '_i$ are the true ages and $\tau_i$ the measured ages. We infer the parameters which are outside of the box, the others are marginalized-out. }
\end{figure}
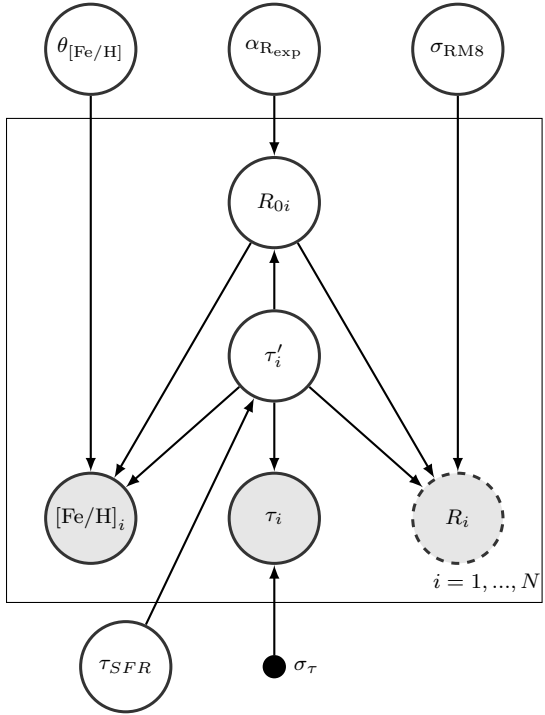
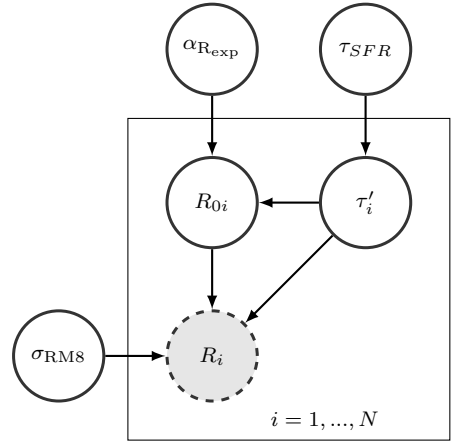
\begin{figure}
\centering
\begin{tikzpicture}
\tikzstyle{main}=[circle, minimum size = 12mm, line width=0.4mm, draw =black!80, node distance = 8mm]
\tikzstyle{connect}=[-latex, thick]
\tikzstyle{box}=[rectangle, draw=black!100]
   \node[main, fill = white!100] (alpha) [ label=center:$\arexp$] { }; 
    \node[main, fill=white!100](tsfr)[right=of alpha, label=center:$\tsfr$]{};
 \node[main, fill = white!100] (R0) [below=8mm of alpha, label=center:$R_{0i}$] { };
\node[main, fill = white!100] (tau) [right=of R0, label=center:$\tau '_i$] { };

  \node[main, fill = black!10, dashed] (R) [below=of R0 ,label=center:$R_i$] {}; 
  \node[main, fill = white!100] (sigma) [left=of R, label=center:$\srm$] { };
  \path
		(sigma) edge [connect] (R)
		(alpha) edge [connect] (R0)
		(tsfr) edge [connect] (tau)
		(tau) edge [connect] (R0)
		(tau) edge [connect] (R)
		(R0) edge [connect] (R);
  \node[rectangle, thick, inner sep=0.mm, fit= (R) (R0) (tau),label=below right:{$ i=1,...,N $}, xshift=-20mm] {};
  \node[rectangle, inner sep=5mm, draw=black!100, fit =  (R)  (R0) (tau)] {};
\end{tikzpicture}

\caption{\label{fig_graphical_model_radius} Sub-model from the model shown in Figure \ref{fig_graphical_model} which is used as the denominator $p(\{ R_i \} ~|~ \ppm)$ in the ratio of probabilities used as the likelihood in this inference (see Eq \ref{eq_general} and the related text). The nomenclature is the same as in Figure \ref{fig_graphical_model}}
\end{figure}
%
%
%

We start using Bayes' rule on
\begin{equation} \label{eq_general}
\begin{split}
p\bigl (\fe , \tau ~|~ R,\ppm\bigr )& = \frac{p(\fe, \tau, R~|~ \ppm)}{p(R~|~\ppm )} \\
&=\frac{\pt \cdot p(\fe, R~|~\tau, \ppm)}{p(R~|~\ppm )}.
\end{split}
\end{equation}
And we will now construct both the numerator and the denominator as two distinct models, summarized respectively in Figures \ref{fig_graphical_model} and \ref{fig_graphical_model_radius}. The numerator is the joint distribution of all data given the model parameters $p(\fe, \tau, R~|~ \ppm)$. But as we do not model the spatial selection function of APOGEE, we should keep the Galactocentric radius $R$ as given, hence the ratio with $p(R~|~\ppm)$.
The first term in the numerator of Eq \ref{eq_general} is the age distribution of red clump stars, given in Eq \ref{eq_age_dist}. The second term in the numerator and the denominator are constructed below. We separate stars younger and older than 8 Gyr in two terms $p_y$ (young) and $p_o$ (old),
as we believe that the model of secular evolution we have laid out is only applicable to 
$\tau < 8$~Gyr. But, in the presence of significant age uncertainties, we must acknowledge the existence of older stars in the Galactic disk without making assumptions on their possible birth radii, enrichment history, and subsequent \rom. For those we specify a less informative metallicity-radius distribution
\begin{equation} 
\label{eq_metallicity_radius_yo}
p(\fe, R~|~\tau, \ppm) = \left\{
                \begin{array}{ll}
                   p_y(\fe, R~|~ \tau, \ppm) ~~ \tau \leq 8 \mathrm{ Gyr} \\
                   p_o(\fe, R~|~ \tau, \ppm) ~~ \tau > 8 \mathrm{ Gyr,}
                \end{array}
             \right.
\end{equation}
where the young term $p_y(\fe, R~|~ \tau, \ppm)$ is derived by marginalizing the joint distribution of metalliticy and birth radii at given time (the age-metallicity-radius relation) $p(\fe, \Ro ~ | ~ \tau, \ppm)$ over stellar birth radii \Ro. Using $p(\fe ~|~ \Ro, R, \tau, \ppm) = p(\fe ~|~ \Ro,\tau, \ppm)$, i.e. the metallicity of stars born at a given birth radius $\Ro$ and time $\tau $ does not depend on their present-day position $R$, we marginalize
\begin{equation}
\label{eq_p_y_feRft}
\begin{split}
p_y(\fe, R~|~\tau, \ppm) &= \int_0 ^\infty  \pRgRot \\
&\times \pRogt \\ 
&\times p(\fe ~|~ \Ro, \tau, \ppm)  \mathrm{d}\Ro,
\end{split}
\end{equation}
with the three terms in the integral being the different aspects of the model. The first two terms are \rom~(Eq \ref{eq:eq_radial_migration}) and the radial birth profile (Eq \ref{eq:eq_radial_birth_profile}), respectively. The third term is the metallicity at birth 
(a Dirac function due to the tight relation Eq \ref{eq:eq_Fe_Ro_t}, or equivalently Eq \ref{eq_Ro_ana}), which we express as a probability distribution function for \Ro : $p(\fe ~|~ \Ro, \tau, \ppm) = \delta (\tRo -\Ro)\cdot \left|\Ro^\prime  \right|$, with ~\tRo~ the analytical solution for the tight relation, defined in Eq \ref{eq_Ro_ana}, and $|\Ro^\prime|$ the Jacobian term relating the distribution in \fe~ and $\tRo (\fe, \tau)$. $\Ro^\prime$ is defined as the inverse of the metallicity gradient 
\begin{equation*}
\Ro^\prime \equiv \frac{\mathrm{d}\Ro}{\mathrm{d}\fe} =  \frac{1}{\dfe} .
\end{equation*}
The Dirac function makes the computation of the integral trivial, simply evaluating the integrand at $\Ro =\tRo$ defined in Eq \ref{eq_Ro_ana}:
\begin{equation}
\begin{split}
p_y(\fe, R~|~\tau, \ppm) & = ~ |\tRo '| ~ \tpRgRot \\ & \times \tpRogt.
\end{split}
\end{equation}
All elements are spelled-out to be recast in $p_y$ of Eq \ref{eq_metallicity_radius_yo}, and we can now do the same exercise with $p_o$.

Guided by the data, we presume that the old term $p_o(\fe, R~|~\tau, \ppm)$ of Eq \ref{eq_metallicity_radius_yo} can be well described by a Gaussian distribution in metallicity and a decreasing exponential in radius. We deem this approximation sufficient for the purpose at hand: this old component is uninformative on \rom~(our interest) and is constructed in order to allow us to treat the large age uncertainties of the data appropriately: with important age uncertainties, we expect a significant number of stars younger than 8 Gyr to have measured ages greater than 8 Gyr, and {\it vice versa}. We define
\begin{equation}
\begin{split}
p_o(\fe, R ~|~ \tau, \ppm) & = p_o(R|\tau, \ppm) \\ & \times p_o(\fe | \tau, \ppm),
\end{split}
\end{equation}
with a radial distribution of old stars (where we keep the variable $\tau $ given, even if there is no explicit dependency, as a reminder that this expression holds given ages greater than 8 Gyr.)
\begin{equation}\label{eq_old_radial_distribution}
p_o(R~|~\tau, \ppm) = \frac{1}{\rold}\exp(-R/\rold)
\end{equation}
with a scale-length $\rold$, and similarly the metallicity distribution common to all old stars, 
\begin{equation}\label{eq_old_fe_distribution}
p_o(\fe ~|~ \tau, \ppm) = \mathcal{N}(\fe, \mfe, \sfe).
\end{equation}
The model parameters part of \ppm ~ here are the old stars scale length $\rold$, their mean metallicity $\mfe$ and their metallicity dispersion $\sfe$. Now, Eq \ref{eq_metallicity_radius_yo} can be fully written and reintegrated into Eq \ref{eq_general}.

Finally, we move on to the denominator in Eq \ref{eq_general}, which is the predicted radial distribution of stars, and can be calculated by  over time:
\begin{equation}\label{eq_radial_distribution}
p(R ~|~ \ppm) = \int _0 ^{\tau_m} p(R ~|~ \tau, \ppm) \pt \mathrm{d}\tau
\end{equation}
with the radial distribution of stars being determined by \rom. Since we presume that conditions at birth are known only for $\tau \leq 8$ Gyr, we separate out older stars again:
\begin{equation}
p(R~|~\tau, \ppm) = \left\{
                \begin{array}{ll}
                   p_y(R~|~\tau, \ppm) ~~ \tau \leq 8 ~\mathrm{Gyr} \\
                   p_o(R~|~\tau, \ppm) ~~ \tau > 8 ~\mathrm{Gyr}.
                \end{array}
              \right.
\end{equation}
The old component $p_o(R~|~\tau, \ppm )$ is the exponential profile introduced above in Eq \ref{eq_old_radial_distribution} with a scale-length $\rold$.
 The radial distribution of $\tau \leq 8 ~\mathrm{Gyr} $ stars is given by the model described in the above subsection. It is determined by the birth radii of stars of age $\tau $, and by their further \rom~after a time $\tau$:
\begin{equation}\label{eq_pR_young}
\begin{split}
& p_y(R~|~\tau, \ppm ) = \\
& \int _0 ^\infty \pRgRot   ~ \pRogt ~ \mathrm{d}\Ro.
\end{split}
\end{equation}
When this expression is inserted back into Equation \ref{eq_radial_distribution}, it leads to a double integral function (extracted in Eq \ref{eq_double_integral}) of the four variables ($R, ~ \tsfr , ~\arexp , ~\srm$). 

The evaluation of such function is computationally expensive: a single evaluation takes about the order of a second, making MCMC sampling  on thousands of stars and tens of thousands of MCMC steps rather slow. We therefore precompute the integral
\begin{equation}\label{eq_double_integral}
\begin{split}
& \int_0 ^{8}\int _0 ^\infty \pRgRot   ~ \pRogt ~ \pt \mathrm{d}\Ro \mathrm{d}\tau
\end{split}
\end{equation}
on a large number of points in the 4D space of $\vec{x} = (R, ~ \tsfr , ~\arexp , ~\srm)$ to interpolate it with precision $0.4 \%$ using a family of highly flexible non linear functions,
\begin{equation} \label{eq:nn}
f(\vec{x}) = \mathbf{W_0} \tanh \Bigl[ \mathbf{W_1} \tanh(\mathbf{W_2} \vec{x} + \mathbf{b_2}) + \mathbf{b_1} \Bigr] + \mathbf{b_0},
\end{equation}
where $\mathbf{W_i}$ and $\mathbf{b_i}$ are matrices of coefficients found by minimizing the difference between Eq \ref{eq_double_integral} and \ref{eq:nn} on the pre-computed points, using a regression gradient descent algorithm.
The interpolation intervals in the parameter space are chosen large enough for our analysis: $3 < R < 15$ kpc, $4 < \tsfr < 14$ Gyr, $0 < \arexp < 1$ and $2 < \srm < 8$ kpc. Possible error propagations during the sum of log likelihood over all data (Eq \ref{eq_overall_likelihood} just below) are discussed in section \ref{section_results}. 

We can now recast all the elements spelled-out above into Eq \ref{eq_general} and build the likelihood function.

The overall likelihood of all the data is given by
\begin{equation}\label{eq_overall_likelihood}
\begin{split}
& \ln{\plik} = \\
&\sum_{i=1}^{N_{data}} ~ \ln{\pliki},
\end{split}
\end{equation}
where \pliki~ is the likelihood of the data on one object, given the model.
Our data \data~ also have uncertainties, dominated by $\tau$ (we neglect those in metallicity and radius as a first approximation) and, therefore, we need to marginalize over these uncertainties. 
\begin{equation} \label{eq:marginalize_errors}
\begin{split}
&\pliki = \\
& \int p_\mathrm{obs}(\tau _i ~|~ \tau)  \cdot p(\fe_i, \tau | R_i) d\tau,
\end{split}
\end{equation}
where $\tau _i$, $\fe _i$ and $R_i$ are the measured age, metallicity and Galactocentric radius (the values in our red clump catalog) and $\tau $ is the potentially true age of the star. Here, $p_\mathrm{obs}(\tau _i ~|~ \tau)$ is the error distribution in age: the probability to measure an age $\tau_i$ given that the possible true stellar age is $\tau$ and measurement uncertainties. This distribution is a Gaussian function in log space, such that for $a = \log_{10}(\tau ) $,  $a_i = \log_{10} (\tau _i)$, $\sigma _{a_i} = 0.2$ dex the error in age, we have $p_\mathrm{obs} (a_i ~|~a, \sigma _{a_i}) = \mathcal{N} (a_i, a, \sigma _{a_i}^2)$. As this noise model may underestimate the errors of very young stars, we apply a different model for stars younger than 0.5 Gyr where errors are Gaussian in linear space with a standard deviation of $\sigma _{\tau_i} = 200 $ Myr,  $p_\mathrm{obs} (\tau_i ~|~\tau, \sigma _{\tau_i}) = \mathcal{N} (\tau_i, \tau, \sigma _{\tau_i}^2)$.
Integral \ref{eq:marginalize_errors} gets evaluated separately for each data point (given each $\ppm$). In practice, we do not need to compute this integral over all the terms in the expression of the distribution $p(\fe_i, \tau | R_i)$ (Eq \ref{eq_general}), but only its numerator because the denominator does not depend on age $\tau$ (hence, we do not propagate the interpolation errors of the term in Eq \ref{eq_double_integral} along this marginalization over age uncertainties).

\subsection{Sampling the Parameter {\it PDF}}

We apply Bayes' theorem on the likelihood function constructed with the analytical disk evolution model described in Section \ref{section_methodology}, and APOGEE red clump giants, to express a posterior probability distribution on the global efficiency of \rom.

The posterior probability distribution on the model parameters is given by
\begin{equation}\label{eq_posterior}
\begin{split}
\posterior (\ppm ~ &|~\data )  = \\
& \frac{ \plik \cdot p_{pr} (\ppm ) }{p_{pr}(\fetau )},
\end{split}
\end{equation}
where we presume $p_{pr} (\data )$, the evidence term that does not depend on the model parameters, to be a constant.
We sample the vector of the 9 free parameters
\ppm $\equiv \{$ \tsfr, \arexp , \rfeh , \tfeh, \dfe,  \srm , \mfe , \sfe, \rold $ \}$,
by means of Equation~\ref{eq_parameterposterior}, and then marginalize over all nuisance parameters $\{$ \tsfr, \arexp , \rfeh , \tfeh, \dfe, \mfe, \sfe, \rold $\}$ to extract a posterior distribution for \rom~$\posterior (\srm ~|~ \data )$. This is done using the MCMC sampler package
Emcee \citep{foreman-mackey_2013}.
In practice, we first perform a maximum likelihood estimation of the parameters using the Nelder-Mead method \citep{NeldMead65}, and sample initial walker positions for 20 Markov chains within small intervals around the best fit results. To compromise the precision of our results and computational time, we perform several fits on different subsets of stars. For each fit, we use a subset of 1500 stars from our low [$\alpha / \mathrm{Fe}$] sample, after having selected further those well in the Galactic disk with $|z| < 1$ kpc. Each chain is sampled with 7000 iterations. We then marginalize over the nuisance parameters to infer the \rom~strength \srm . Our prior on
$\srm$ is set by the restricted space where the interpolation of equation \ref{eq_double_integral} is valid: $2 < \srm < 8$ kpc. The priors on other model parameters are also flat, we only constrain distances and durations to be positive.

%
%
%
%
%
%

\begin{figure*}
\includegraphics[scale=0.5]{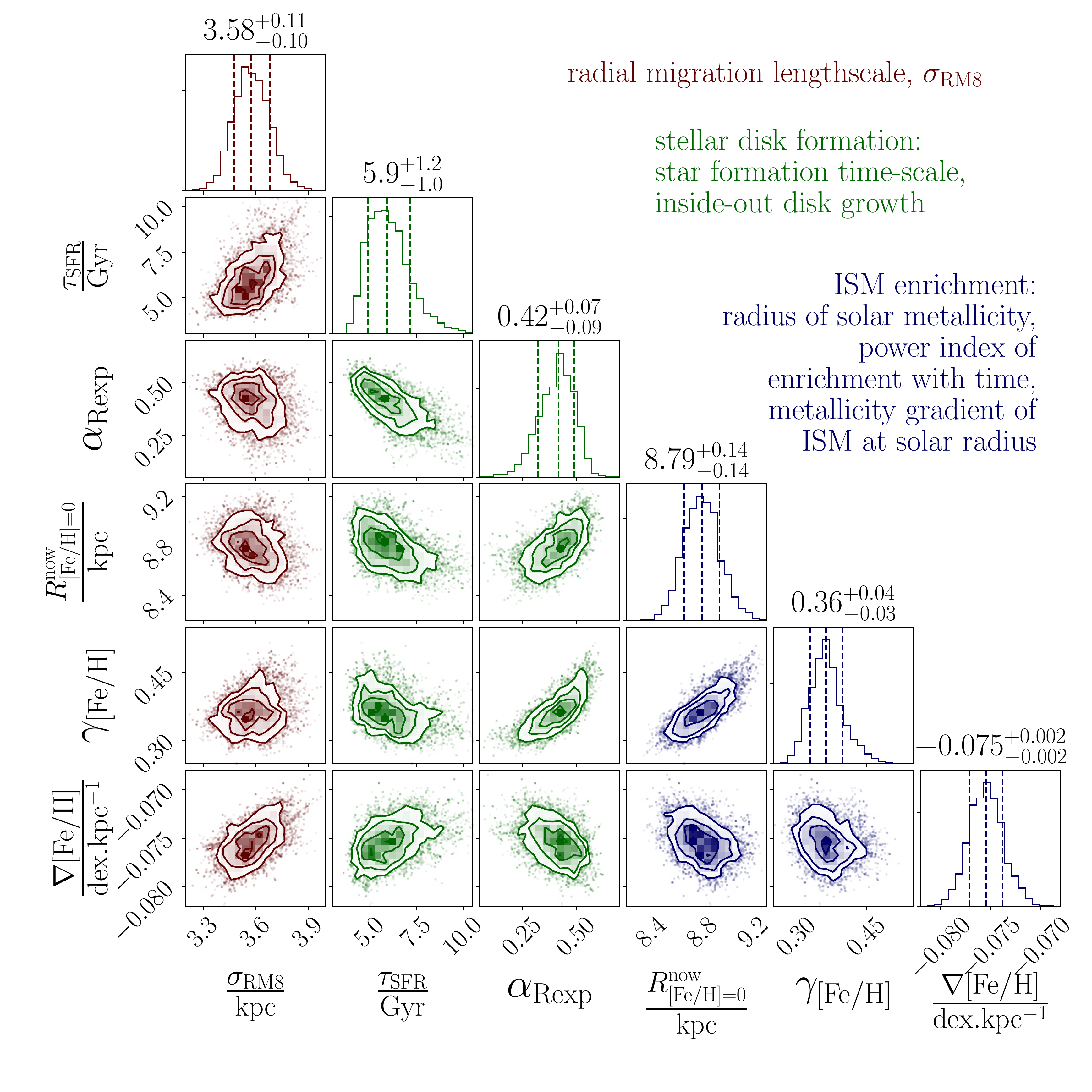}
\caption{\label{fig_corner} Posterior distribution of the 6 parameters of the Galactic disk evolution model. From left to right: the main parameter of interest for \rom~strength \srm ~ in kpc (Eq \ref{eq:eq_radial_migration}), followed by the nuisance parameters: star formation time-scale \tsfr ~in Gyr (Eq \ref{eq_age_dist}), the parameter characterizing inside-out disk growth \arexp ~(the Miky Way disk was approximately 40\% smaller at its possible formation 8 Gyr ago, Eq \ref{eq:eq_radial_birth_profile}). Then come the three parameters characterizing the enrichment of the interstellar medium (ISM)  as a function of time and galactic radius (Eq \ref{eq:eq_Fe_Ro_t}): the radius where the ISM metallicity is solar \rfeh ~in kpc, the power index characterizing the gradual chemical enrichment of the ISM with time $\gfe$, the metallicity gradient of the ISM at the solar radius $\dfe$ in dex $\mathrm{kpc^{-1}}$. 
A complete version of the posterior in the 9D parameter space (that includes the parameters of the less informative component for old stars) is in appendix.
}
\end{figure*}

\section{Results}\label{section_results}

We now summarize the results obtained from fitting our disk evolution model to the low-$\alpha$ APOGEE red clump data, described in Section~\ref{section_data}. The maximum likelihood estimates (Eq \ref{eq_overall_likelihood}) for the model parameters are presented in Table \ref{table_best_fit}. All 20 chains of the MCMC converged with 7000 iterations on subsets of 1500 stars out of the 17,500 low-$\alpha$ available stars of the sample. We show the posterior distributions for the parameters of immediate interest in Figure \ref{fig_corner}; it shows that all parameters are well constrained by the data, with some covariances but no degeneracies. The full version of the figure, that shows the exploration of the whole parameter space including all nuisance parameters, can be found in Figure \ref{fig_corner_full} in the Appendix. 

We first focus on quantifying on \rom, show the model calculation for the best fit parameters, and then comment briefly on the other parameters.

\begin{table}
\vspace{0.25cm}
\caption{Best fit MLE parameters\label{table_best_fit}}
\begin{tabular}{lcr} 
\hline
\ppm & 			Best fit & Description \\
\hline
\tsfr /Gyr& 			6.8		& Star formation time-scale\\
\arexp & 				0.3	& Inside-out growth \\
\rfeh /kpc&				8.7 	& R($\fe = 0,\tau=0$)\\
$\gamma $ &				0.3		& Enrichment power index \\
$\dfe$ /dex kpc$^{-1}$&	-0.075	& $\fe _\mathrm{ISM}$ gradient \\
\srm /kpc & 			3.5		& \srom~ strength \\
$\rold$ /kpc & 			2.5		& Scale-length old disk \\
$\mfe $/dex &		-0.05	& Mean metallicity $\tau > 8 $ Gyr\\
$\sfe $/dex &		0.15 	& std metallicity $\tau > 8 $ Gyr\\
\hline
\end{tabular}
\vspace{0.25cm}
\end{table}

\subsection{Radial orbit migration}
\begin{figure}
\includegraphics[scale=0.55]{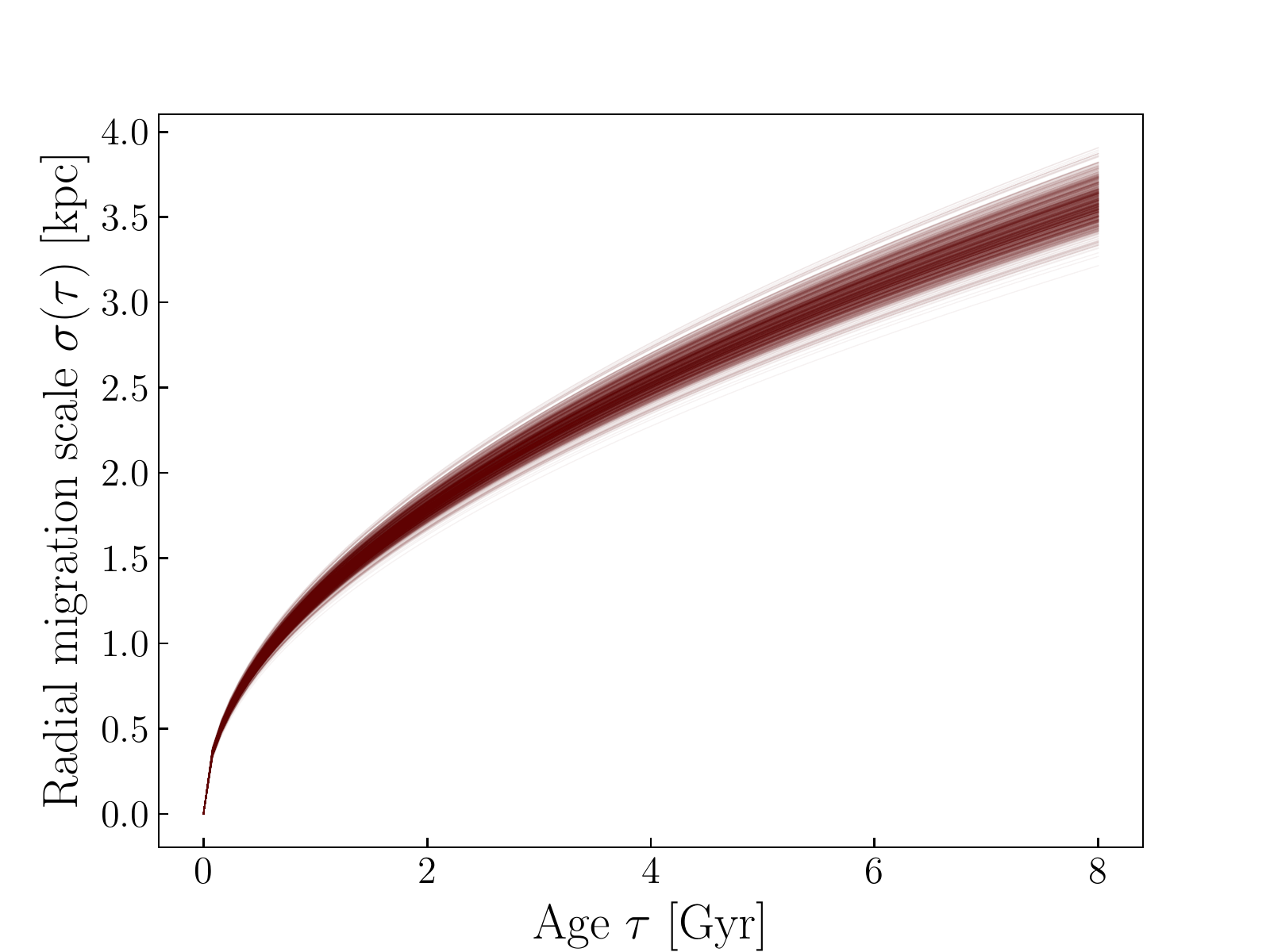}
\caption{\label{fig_result_radial_migration}Radial orbit migration strength inferred in this study with respect to stellar ages.}
\end{figure}

Fig \ref{fig_corner} shows that in this modelling context, the strength of \rom~is very well constrained. Marginalizing the posterior distribution over the nuisance parameters gives an estimate of \srm ~of about $3.6 \pm 0.1$ kpc (see Figure \ref{fig_corner}). This represents the length-scale over which the oldest stars (8 Gyr) have spread around their birth radii. In Fig~\ref{fig_result_radial_migration}, our \rom~estimate $\sigma (\tau)= 3.6~\mathrm{kpc} \sqrt{\tau /~8\mathrm{Gyr}}$ is illustrated by sampling from posterior distribution, i.e. the MCMC chains in Figure \ref{fig_corner}). This result quantifies that the present-day radius is a poor proxy for the birth radius, compared to the metallicity at given age.

Mathematically, \srm~ quantifies the distance between the present Galactocentric radius of a star and the birth radius {\it expected} from the global model fit. Whenever the quantity of interest is a ``scatter'' one must explore to which extent it is attributable to other model shortcomings.
We have therefore explored model variants and have found that this \rom~strength estimate is rather robust. We exercised MCMC estimates holding other model parameters fixed to diverse values, leaving the \rom~strength estimate robust.


\subsection{Other parameters}
But along with the \rom~strength, the model also constrains all other aspects: star formation history, inside-out growth, and the enrichment history of the Galactic disk. While these are mere nuisance parameters when constraining \rom~strength, they are informative about the Galactic disk evolution over the last 8~Gyr.

\paragraph{Star formation history: \tsfr}

\begin{figure}
\includegraphics[scale=0.55]{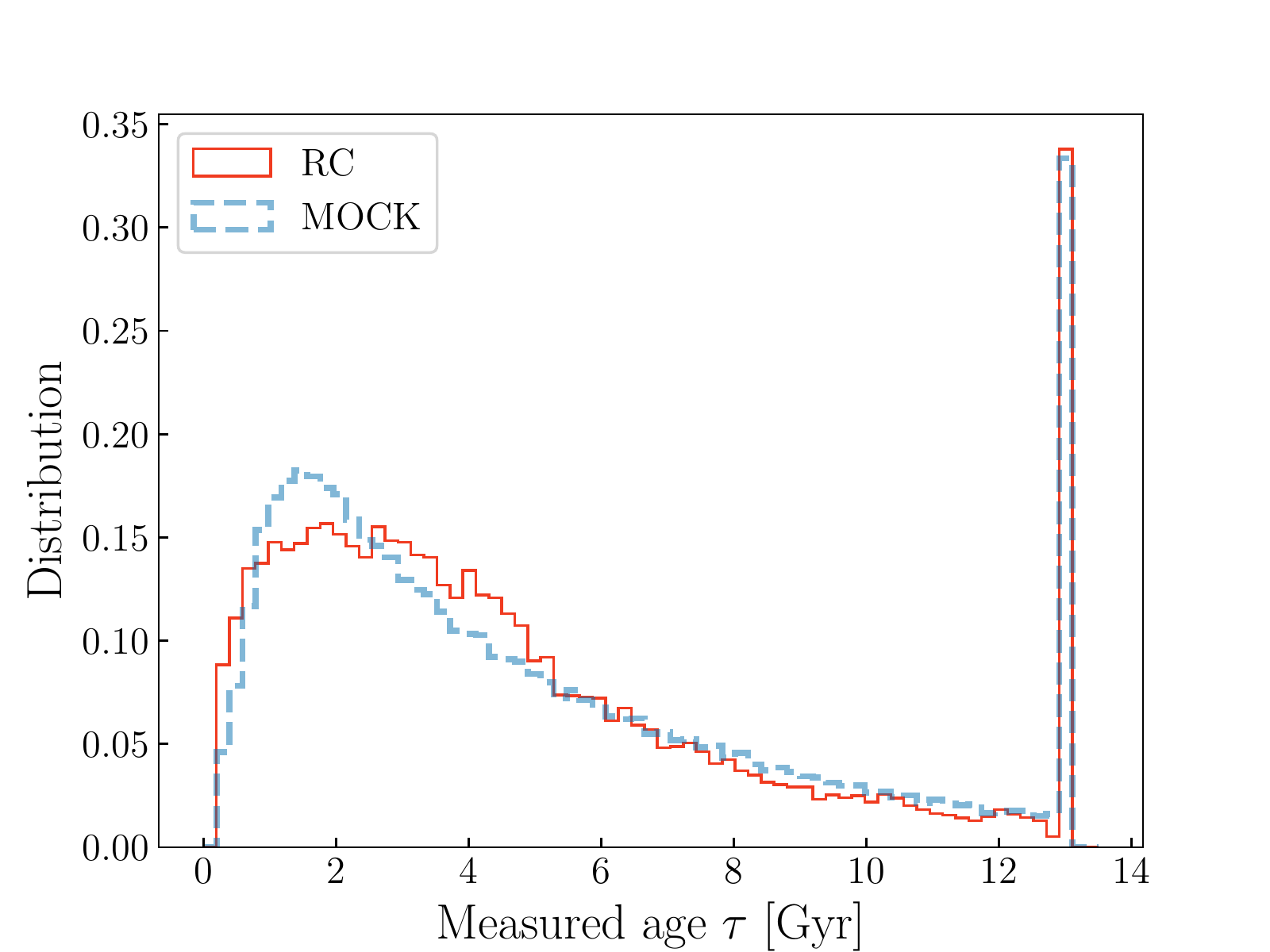}
\caption{\label{fig_age_dist_res}Age distribution of the low-$\alpha$ APOGEE red clump stars sample: observed (red) and predicted (blue) by the model evaluated at the best MCMC parameters and with reproduction of the effects of the spatial selection function (subsection \ref{sub_sec_test_verif}). The peak of stars at 13.4 kpc are stars with initially measured age greater than the age of the Universe, of which the age was brought back to this exact value, see \cite{ness_etal_2016}.}
\end{figure}

The data favor a star formation time-scale $ \tsfr = 6 \pm 1$ Gyr for the Galactic low $[\alpha / \mathrm{Fe}]$ disk. This value seems rather low given prior expectations of an extended star formation in the thin disk. We find that this estimate depends strongly on the assumed form of the age distribution at young ages ($< 1$ Gyr); and it is sensitive to the details of the selection: e.g., we find a larger star formation time-scale if we select $|z|<1$ kpc stars rather than if we select $|z|<1.5$ kpc; this should be expected as the proportion of young stars is larger near the mid-plane. Given the age distribution varies with Galactocentric radius and height above the plane, the uneven APOGEE pointings could induce some $\tsfr$ bias for which we do not correct. Additionally, this estimate is degenerate with the old stars scale-length parameter \rold~(Figure \ref{fig_corner_full} in Appendix). This is due to the spatial selection function limited to 5 kpc from the center of the disk: predicting a fast star formation (many old stars) with a small scale-length is, according to this model, roughly equivalent to predicting a slow star formation (less old stars) but more extended in the disk, preserving the overall observed ratio of young to old stars (the Galactocentric radius ranges are 5-14 kpc: we do not see an old stellar population when it is well concentrated in the inner disk). This is because these two scenario will predict the same amount of old stars in the \textit{observed} regions of the Galactic disk. However, even if our estimate of \tsfr~is questionable, we note that (1) this does not seem to affect our \rom~strength estimate, and (2) the observed age distribution of red clump stars is well reproduced, as illustrated in Figure \ref{fig_age_dist_res} which shows a comparison of the observed age distribution of red clump stars to the one predicted by the model (the details of this procedure are described in subsection \ref{sec_res_compare_data}).

\paragraph{Inside-out growth: \arexp}
The growth (i.e. star formation) of the Galactic disk was modelled by the scale-length parameter $R_\mathrm{exp} (\tau) = 3 \mathrm{kpc} (1 - \arexp \frac{\tau}{8 \mathrm{Gyr}})$ of newborn stars. We find $ \arexp = 0.42 \pm 0.09$. This implies that the disk was about 40 \% smaller 8 Gyr ago. This is consistent with observations of high redshift disk galaxies \citep[e.g][]{van_dokkum_etal_2013}. However, we report that the estimate for this parameter was very sensitive to the assumed functional form for the metallicity profile combined with the age distribution, with covariances with \tsfr.

\paragraph{Metallicity profile and enrichment history: \{\rfeh, \dfe, \gfe\} }
The metallicity profile of the cold gas in the disk is described in our model by a simple straight line in radius with a negative gradient. The two model parameters that characterize the metallicity profile are \rfeh : the Galactocentric radius at which the star-forming gas metallicity is solar, corresponding to an arbitrary zero point, and \dfe : the present-day metallicity gradient at \rfeh. As these are two "present-day" properties, the youngest stars of our sample are expected to provide the strongest constraints on these parameters. We find the radius of solar metallicity to be about $\rfeh = 8.8 \pm 0.2$ kpc. The metallicity gradient \dfe~is found to be $-0.075 \pm 0.002 \mathrm{dex.kpc^{-1}}$. The values of these two parameters are consistent with the left panel of Figure \ref{fig_data_metallicity_radius_plane}, for which we plot the metallicity profile of the young red clump stars. The densest region for \fe = 0 dex is close to 8kpc. We note that \cite{sanders_binney_2015} find different results with their model on the Geneva-Copenhagen Survey data \citep{nordstrom_etal_2004}, with the radius of solar metallicity of 7.37 kpc and a shallower metallicity gradient of -0.064 $\mathrm{dex} \cdot \mathrm{kpc^{-1}}$, and \cite{genovali_etal_2014} measure a gradient of $-0.060 \pm 0.002 \mathrm{dex} \cdot \mathrm{kpc^{-1}}$. More recently, \cite{anders_etal_2017} measured the stellar metallicity gradients for red giants in different stellar age bins, and found about $-0.058 \pm 0.008 \mathrm{dex} \cdot \mathrm{kpc^{-1}}$ for stars younger than 1 Gyr.

\begin{figure*}
\includegraphics[scale=0.6]{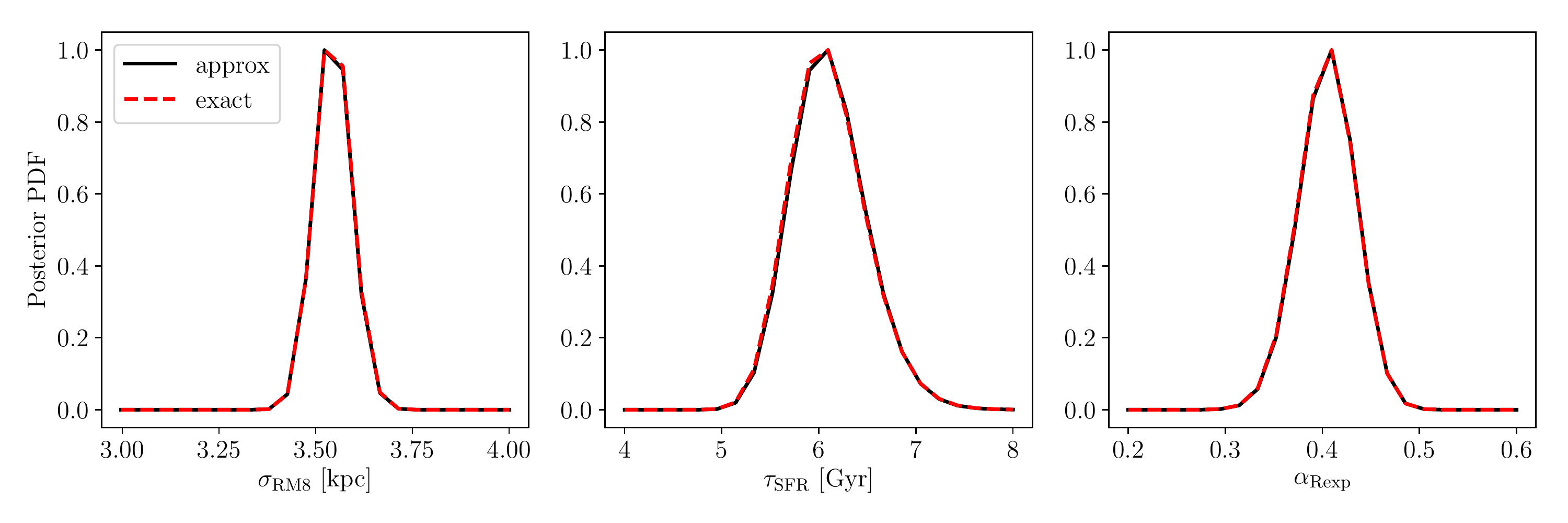}
\caption{\label{fig_result_test_NN} Comparison between the posteriors that contain an approximated term in black (for computational optimization purpose) and the exact posteriors in dashed red (not normalized). We take cuts in the posterior probability density in the three dimensions that were approximated in Eq \ref{eq_double_integral} and compute them on 22 points for each panel. From left to right: \srm, ~  \tsfr, ~ \arexp. All other parameters in \ppm were taken at their best MCMC value. The black and red lines overlap well: the interpolation errors, even after a sum of log likelihoods over 1500 stars, propagate slowly and have no effect on where the posterior maximum is located, nor on the width of the distribution. This gives us confidence that the 'best MCMC' parameters and their corresponding errors were evaluated well enough, without dramatic effects from the interpolation of Eq \ref{eq_double_integral}.}
\end{figure*}

The enrichment history at any radius of the disk is described in our model by a power law of time with index \gfe. The best MCMC value is $\gfe = 0.36 \pm 0.04 $. The metallicity of the interstellar medium is plotted with respect to look-back time in Figure \ref{fig_age_metallicity} (using the MLE results). This result is different from (semi) analytic models used previously in the literature \citep{schonrich_binney_2009a,schonrich_binney_2009b,sanders_binney_2015}, where the enrichment of the interstellar medium generally increases faster at early times and is almost flat at late epochs. Here, we find that the gas metallicity grows continuously at all radii up to the present day. This is, however, very much consistent with Milky Way-like simulations \citep{grand_etal_2018}, and further tests with different metallicity profile forms would be interesting.

\paragraph{The nuisance model for the disk before 8 Gyr ago}
We built a less-informative ``nuisance'' model  the Milky Way disk 
older than 8~Gyr, to avoid sharp age cuts. But these stars enclose information on the star formation history of the Milky way: in essence, they help to constrain the \tsfr~ parameter only. The other three model parameters that correspond to our old stars model are \mfe, \sfe, \rold. The mean and variance of the old stars metallicity appear to be robust estimates and do not show degeneracies with other parameters. The MCMC exploration shows \rold  to be about 2.4 kpc (see the full corner plot in appendix, Figure \ref{fig_corner_full}). This value is physically coherent with our prior knowledge on the disk. We note that this \rold ~parameter, which we model as the ``old disk scale-length", is degenerate with \tsfr. This is not surprising for two reasons, one is physical, the other is a model caveat: it was shown in \cite{bovy_etal_2012} that the scale-length of the stellar disk was a function of their age (in a chemical sense, [$\alpha /\mathrm{Fe}$]), and being able to over or under predict the number of old stars in regions not covered by APOGEE (e.g., the inner disk $< 5$ kpc) could be a caveat of this model. 

\subsection{Tests and verifications \label{sub_sec_test_verif}}

In this section, we examine some of the model and methodological shortcomings or restrictions that could bias our inferences, such as the approximations made to minimize the computational cost of likelihood evaluations and the convergence of the MCMC. We further address the robustness of the \rom~strength estimate. Finally, we confront the predictions of our model evaluated at the best MCMC values in the space of the data.

\subsubsection{Technical verifications}
The term $p(R~|~\ppm)$ in likelihood function (Eq \ref{eq_double_integral}, represented in Figure \ref{fig_graphical_model_radius}), was interpolated. Interpolation errors on a set of 20,000 test points are less than 0.4\%. To see whether interpolation errors have propagated during the overall product of the likelihood over the 1500 stars used for inference, we choose slices in the parameter space at the best MCMC values, and compare the (expensive) true likelihood evaluations to the approximated values. The relative differences are small, as can be seen in Figure \ref{fig_result_test_NN} that shows three slices of the posterior distribution. Additionally, the generation of mock data and the model itself $p(\fe, R, \tau | \ppm)$ do not rely on the approximation in $p(\fe, \tau |R, \ppm)$, so comparisons between the model prediction and the data in Section \ref{sec_res_compare_data} also give us confidence that the posterior was approximated well enough for our purpose. 

To address the question of the MCMC convergence and the exploration of the parameter space, we have run several more MCMC chains, where the walkers were started in more extended ranges than just the MLE neighborhood. The results remained close to those presented in Fig \ref{fig_corner}. We have also performed MCMC on three different random batches of 1500 stars. We found that \rom~strength \srm, the present-day cold gas metallicity gradient at solar radius \dfe, and the radius of solar metallicity in the interstellar medium \rfeh~ were extremely well constrained. However, the star formation time-scale \tsfr~ showed some variability (the best \tsfr~ varied between 5 Gyr and 7.5 Gyr) depending on the sets of stars used, but as discussed in the subsection above, this parameter showed to be sensitive to biases and has ~ 1 Gyr uncertainty. Finally, we calculated the potential scale reduction factor, estimating the ratio of variances within single chains and between several chains to about $< 1.03$ and the autocorrelation time, to 180 steps (where the MCMC ran 7000 iterations).

\begin{figure*}
\includegraphics[scale=0.55]{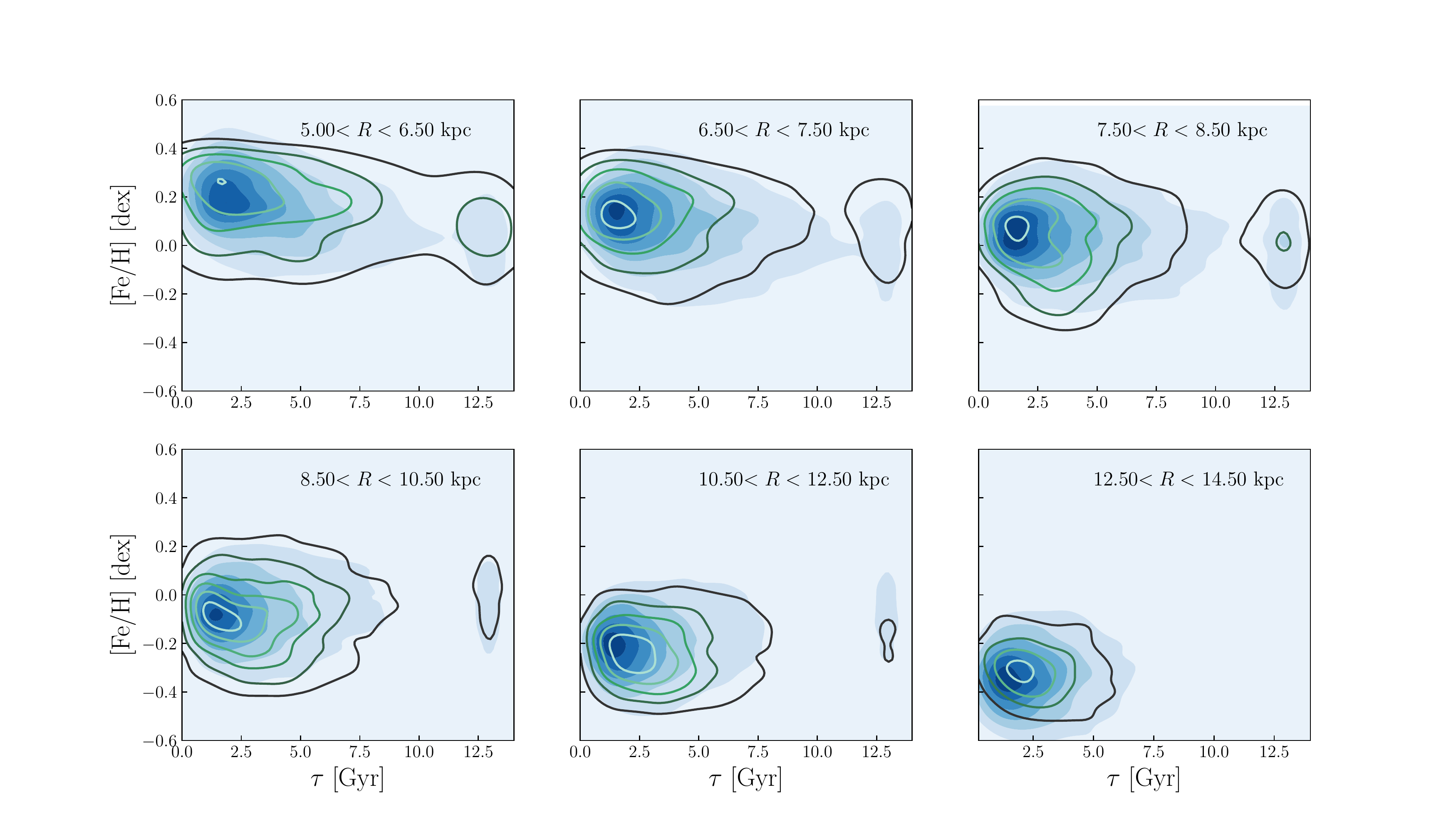}
\caption{\label{fig_result_age_metallicity_data_model}Density of red clump stars in the age--metallicity plane at six Galactocentric radii: 6, 7, 8, 9, 11 and 13 kpc. The model predictions using the best MCMC values are represented by the shaded background. Isolines on the foreground represent smoothed isocontours of \fe --$\tau$ distributions of red clump stars data in the different radial bins. The scatter in age--metallicity is well reproduced through the effect of \rom. The over-density (dark area) predicted by the model (background) for ages near 2 Gyr is the effect of the selection of red clump population (see Figure \ref{fig_age_distribution} for their age distribution that peaks near 2 Gyr).\vspace*{6mm}}
\end{figure*}


\begin{figure*}
\includegraphics[scale=0.55]{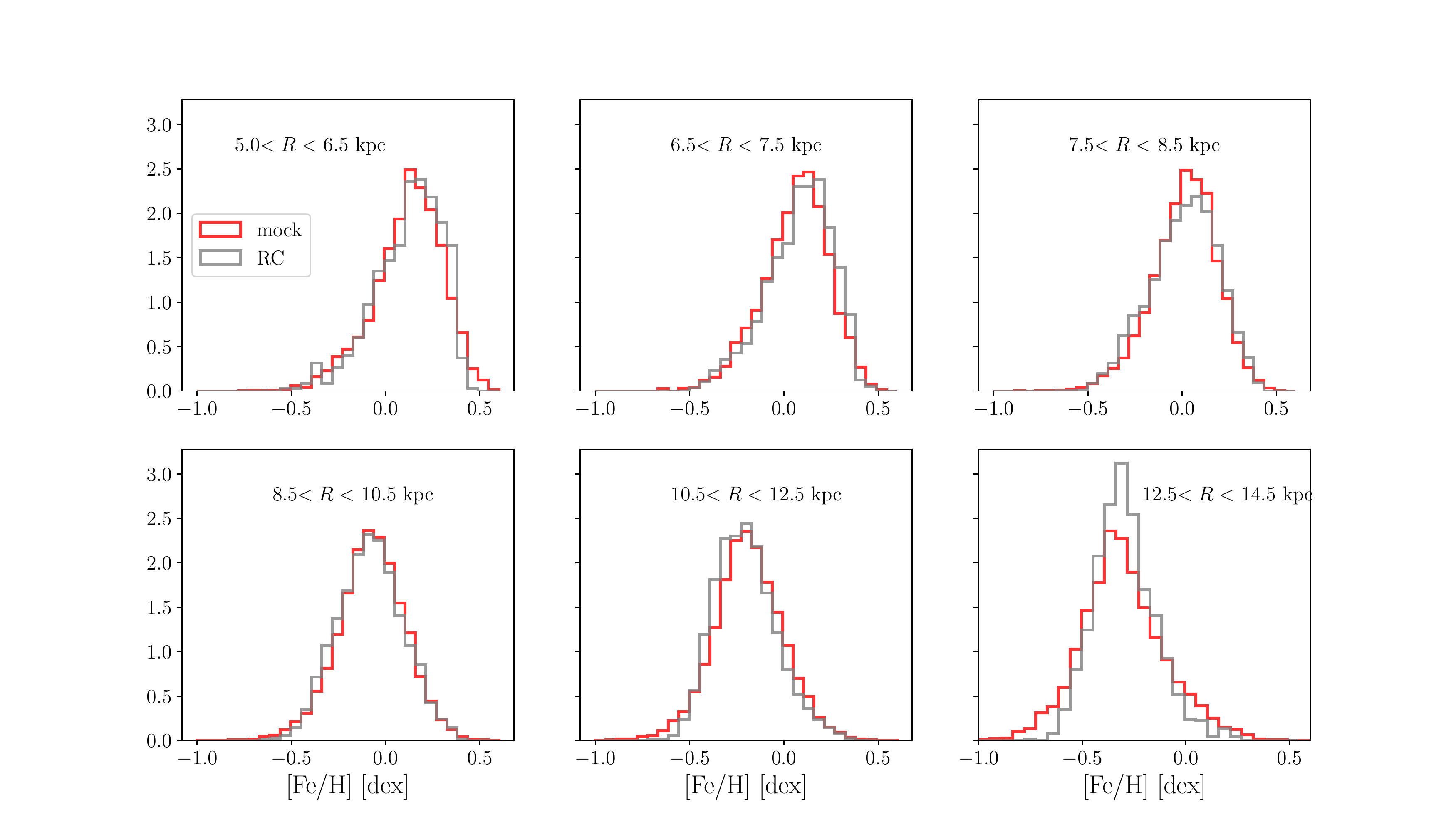}
\caption{\label{fig_result_MDFs_data_model}Metallicity distribution functions (MDFs) at six Galactocentric radii: 6, 7, 8, 9, 11 and 13 kpc. The model predictions using the best MCMC values are represented by the red histograms. Red clump stars MDFs are shown in grey. The metallicity distribution functions are well reproduced and a metallicity gradient is visible (shift of the MDFs maxima as $R$ changes), except for the outer disk where the data show the limitations due to the rigidity of our model.\vspace*{3mm}}
\end{figure*}

\subsubsection{Model predictions in the data space \label{sec_res_compare_data}}

We generated a mock data set to compare with APOGEE red clump sample, using rejection sampling on the different aspects of the model evaluated at the best MCMC values. For comparison with APOGEE data, we reproduced the age distribution of red clump stars using the same functional form as in \citep{bovy_etal_2014}, and introduced some scatter for the age uncertainties using our noise model (a Gaussian of width 0.2 dex in $\log_{10}$ age, and a floor of $\sigma _\tau =$ 200 Myr uncertainties for stars younger than 0.5 Gyr). We imitated the possible effect of the radial selection function in our data set using importance sampling (thereby reproducing the radial distribution of stars in our data set). This is a relevant test to do, as inference of the parameters was performed only on a small fraction of the overall catalog: the MCMC was performed (multiple times) on 1500 low-$\alpha$ stars randomly selected in the red clump catalog. Asking if the model can describe the rest of the 17,500 stars is therefore an interesting test.
We show the results in three different plots allowing data comparison. First, we map the age--metallicity plane $p(\fe, \tau | R)$ with contours of both our mock data and the APOGEE red clump sample in different radial bins, see Figure \ref{fig_result_age_metallicity_data_model}. The observed trends are well reproduced in the Solar neighborhood and inner disk, but the last panel (13 kpc) shows that the model predicts a distribution broader (in metallicity, so in the vertical direction in Figure \ref{fig_result_age_metallicity_data_model}) than the observed distribution. We suspect that the main differences between our predictions and the data come either from our restrictive model for the evolution of metallicity gradient, or from the fact that \rom~strength could depend on radius whereas we fitted a global value. The effect of the metallicity profile will be investigated in the subsection \ref{subsection_model_variant}.

Secondly, we integrate the age--metallicity plane $p(\fe, \tau ~|~ R)$ with respect to age to show the metallicity distribution functions at given radii $p(\fe ~|~ R)$. These distribution functions are well reproduced in most of the disk, showing the expected positive skewness appearing due to \rom~\citep{hayden_2015, loebman_etal_2016, toyouchi_chiba_2018}. The difference between observed and predicted metallicity distribution function at 13 kpc is more obvious here.

\begin{figure*}
\includegraphics[scale=0.75]{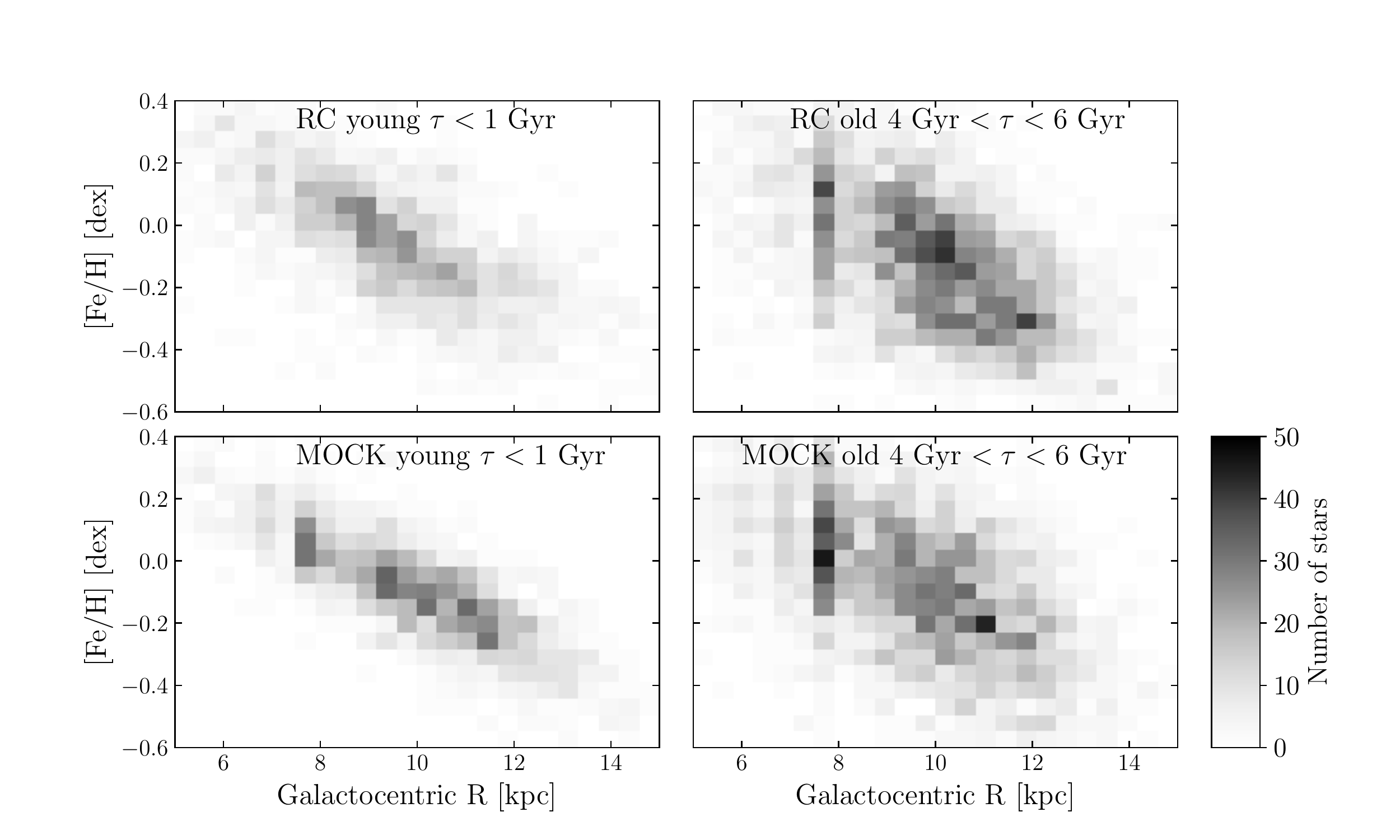}
\caption{\label{fig_res_metallicity_profile_exp}Number density distribution of red clump (RC) stars (top) and mock data (bottom) in the plane of metallicity ([Fe/H]) and Galactocentric radius, for two ages bins: young stars (less than 1 Gyr, left), and older stars (measured age between 4 and 6 Gyr, right). The total number of mock stars equals the total number of red clump stars.}
\end{figure*}

Finally, we compare the prediction in the radius -- metallicity plane; in Fig \ref{fig_res_metallicity_profile_exp}, where the radial spread at fixed metallicity clearly increases with age at similar rates both for the observed and mock data; and the overall metallicity gradient and the broadening of distributions with time seems to be well reproduced.
At young ages, the spread of radii at given metallicity is slightly underestimated by the model. This is because (1) our model assigns any metallicty scatter at given radius and age to \srom, and at young ages, the probability distribution of a star tends to a Dirac function (Eq \ref{eq:eq_radial_migration}) and (2) we neglected measurement errors in metallicity. We note that if star clusters are intrinsically homogeneous but data show additional metallicity spread at young ages for a given Galactocentric radius, azimuthal variations of metallicity could be probed by adding one more parameter accounting for scatter in the metallicity-radius-age relation. 

\subsubsection{Model variant \label{subsection_model_variant}} 

The tests presented above showed that (1) the fitting procedure went well for most parameters and the model describes the observations well for most of the Galactic disk, (2) the estimate of \rom~strength is robust, but (3) the model does not reproduce the outer disk observed metallicity distribution functions. This can be interpreted in several ways:
(a) The metallicity profile functional form that we assumed is rigid: it describes a straight line for which we fit the gradient \dfe, the zero point (after translation of \rfeh ) and the time evolution of the zero point \gfe. But the gradient itself could evolve with time, as pointed out in \cite{minchev_2018}. Or the assumption that the metallicity profile is well described by straight line could be a too simple extrapolation of the observed gradients. \cite{sanders_binney_2015} used a different functional form describing a decreasing exponential in radius:
\begin{equation}
\fe = F_m \Bigl( 1 - \exp(-\frac{-\dfe(R - \rfeh)}{F_m} ) f(\tau)\footnote{with a different enrichment prescription $f(\tau)$, which we also tested separately}    \Bigr)
\end{equation}
We tested this form with several MCMC procedures, and the estimate of \rom~with this model was $\srm = 4.0 \pm 0.1 $ kpc, which remains close to our current result and confirms the robustness of the estimate of \srm~in the present study. Additionally, the outer disk was very well described by mock data from a fit to this model. However, the model predictions in the inner disk were problematic: we systematically overestimated the metallicity of stars born in the inner disk. \cite{sanders_binney_2015} reported the same high metallicity trend while modelling the Solar neighborhood. This gives us confidence that the metallicity profile description is a key ingredient in such modelling, and any model-induced rigidity can affect the results significantly (here: reproducing the metallicity distribution functions, even though the estimate of \rom~strength was affected by less than 15\%).
(b) Another interpretation for the disagreement between model predictions and observed metallicity distribution at 13 kpc could be that \rom~occurs differently at different strengths at different radii. We note we used only one global parameter to describe \rom~over the whole disk, and that outer disk stars are not well described by our global fit, suggesting that a \srm~is a spatial average of a Galactocentric radius-dependent \rom~strength.

%
%
%
%
%
%
\section{Discussion and conclusions}
\label{section_ccl}
\subsection{Summary and implications}
In this study we have quantified the global efficiency of \rom~in the Galactic disk. We have built an analytical disk evolution model, in good part inspired by \cite{sanders_binney_2015},
that combines the distribution of star formation in radius and time with the chemical enrichment of the ISM, and with subsequent diffusive migration of the stars' orbital radii. Our model does not attempt to differentiate whether changes in the orbital radius are to be attributed to churning or blurring.

We have applied this to a set APOGEE red clump stars with age estimate, a large sample of stars with precise distances (covering 5~kpc$\lesssim R \lesssim 14$~kpc) and metallicities; this is the first time that such a large and radially extensive data set with consistent estimates of \fe and $\tau$ has been available. We sidestepped the complex spatial selection function of this survey and accounted for the 0.2~dex age uncertainties. 

This has enabled for the first time an estimate of the overall \rom~efficiency throughout the Galaxy, using $\{R,\fe ,\tau\}_i$. Previous studies of radial migration focused on the Solar neighborhood \citep[Geneva-Copenhagen Survey data]{sanders_binney_2015}. Other studies
of large radial extent in the Galactic disk, using e.g., APOGEE, had focused  mainly on recovering the present day stellar metallicity distribution functions without the explicit use of stellar ages
\citep{hayden_2015, toyouchi_chiba_2018}. The model draws its constraints from the mean metallicities at each age and (present-day) radius, and from the spread of these metallicities (growing with age).

Our basic result is that APOGEE data tell us quite directly in this modelling context that \rom~in the Galactic (low-$\alpha$) disk is strong, 
$\langle \bigl | R(\tau)-R_0\bigr |\rangle \approx 3.6 ~\mathrm{kpc} \sqrt{\tau / \mathrm{8~ Gyr}}$. This means that the characteristic distance over which stars migrate over the age of the disk is comparable to the half-mass radius of the Milky Way disk. Qualitatively, this has of course been implied by a number of earlier studies, \citep[e.g.,][]{schonrich_binney_2009a}; and it has been implied by numerical studies of disk dynamics \citep[e.g.,][]{roskar_etal_2008b,minchev_famaey_2010}. But a stringent and global modelling-based estimate of this efficiency from stellar data across the Galactic disk had not been explored before.

This result has a number of astrophysical implications. First, it tells us that for disk stars older than a few billion years, the current radius is not a particularly good indicator of the stars' birth radii. The combination of age and metallicity should be a better predictor of $R_0$. 

For example, the Sun's age (4.6~Gyr) and [Fe/H]$\equiv 0$ implies in our model context that it was born at $5.2\pm 0.3$~kpc: it has migrated outward by about 3 kpc since. While it is true that [Fe/H]$_\mathrm{now}(R_\odot)\approx 0$ dex,
the continuous ISM enrichment at all radii does not imply $R_\odot\approx R_\mathrm{birth}$. This quite precise $R_\mathrm{birth}$(Sun) estimate is 
in agreement with the broad prediction from chemo-dynamical simulations of \cite{minchev_etal_2013} (between 4.4 and 7.7 kpc), but has 2 kpc difference with the recent results of \cite{minchev_2018} (7.3 $\pm$ 0.6 kpc). It is in contradiction with Solar birth location predictions based on backward integration of \cite{martinez-barbosa_etal_2015}, finding that the Sun should come from the outer disk rather than from the inner disk. 

Further, our results show and confirm that -- even in the absence of any significant violent relaxation in the last $\sim 8$~Gyr -- the stellar distribution in the Galactic disk experiences significant ``dynamical memory loss''; the angular momentum of stars in the disk is not even approximately conserved, though many of these stars may now still be on near-circular orbits.  The value of \srm , when combined with the radial velocity dispersion of the disk, implies that churning is a considerably stronger effect than blurring in the Galactic disk.

We derived these results without having to drawing on detailed chemical tagging \citep{freeman_bland-hawthorn_2002}. Instead, we relied on the assumption that the spread in birth metallicities among stars born at the same time at the same radius was small over the last 8~Gyr; this is in some sense the most elementary version of chemical tagging.

To the extent that our Galactic disk is typical for large disk galaxies
\citep{rix_bovy_2013,bland-hawthorn_gerhard_2016}, this result helps explain 
why the stellar mass density profiles of disks are smooth and approximately exponential. \citet{herpich_etal_2017} have shown that asymptotically efficient radial migration leads to exponential profiles. Of course, this ``thermodynamic limit'' of maximal angular momentum entropy would erase all abundance gradients, in conflict with observations.  
Our analysis here shows that strong \rom~may happen, and still match the radial abundance gradients (at least [Fe/H]) in detail.


\subsection{Current limitations and future prospects}
In concluding, it may be good to recall some of our main model assumptions and simplifications:  (1) We used a restricted \rom~description, assuming it
to be constant across the disk and with a specific time dependency. While it is plausible that \rom~occurs over wide range of radii and over much of the disks' evolution history, it would be good to explore whether the extensive orbit-abundance data sets allow to constrain the presumably more complex radial or temporal dependence of \rom~efficiency \citep{brunetti_etal_2011,kubryk_etal_2013,toyouchi_chiba_2018}. (2) At a basic level, our model explained ``scatter'' in data with \rom . This always begs the question whether other sources of scatter have been considered exhaustively. For example, we treated the (dominant) age uncertainties by explicit marginalization in the model, but did not do the same for [Fe/H] uncertainties to save computational expense. Also, future work could generalize the assumption that the abundance scatter at a given birth radius and epoch was zero, to the assumption that it was merely ``small''.  And, (3) we restricted our \rom~ analysis to modelling of Galactocentric radius, while angular momentum and radial action should be modelled to best differentiate churning from blurring. The arrival of data from Gaia DR2 \citep{GDR2} suggests such a generalized analysis as the next step.

We also eliminated the explicit $R_i$-dependence of the model, to eliminate the model's dependence on the detailed spatial selection function. But this approach ``to ignore the observed radius distribution'' also eliminates much valuable information. Future modelling could tackle the spatial selection function head-on \citep[e.g.][]{bovy_rix_etal_2016}.

Finally, we have only considered [Fe/H] in this work. The vast stellar data sets of more detailed element abundance measurements must contain much information about where stars were born and how much they migrated. This, too, bears detailed modelling.

\section*{Acknowledgements}
We thank Wilma Trick, Coryn Bailer-Jones and Morgan Fouesneau for valuable conversations.
N.F. acknowledges support from the International Max Planck Research School for Astronomy and Cosmic Physics at the University of Heidelberg (IMPRS-HD). H.-W.R. received support from the European Research
Council under the European Union's Seventh Framework
Programme (FP 7) ERC Grant Agreement n. [321035].

The following softwares were used during this research: Astropy \citep{astropy}, Matplotlib \citep{matplotlib}, Pytorch \citep{pytorch}, Emcee \citep{foreman-mackey_2013}.
Figures \ref{fig_corner} and  \ref{fig_corner_full} were produced using the package Corner \citep{corner}.

%
%
%
%
%
%

\begin{figure*}
\includegraphics[scale=0.35]{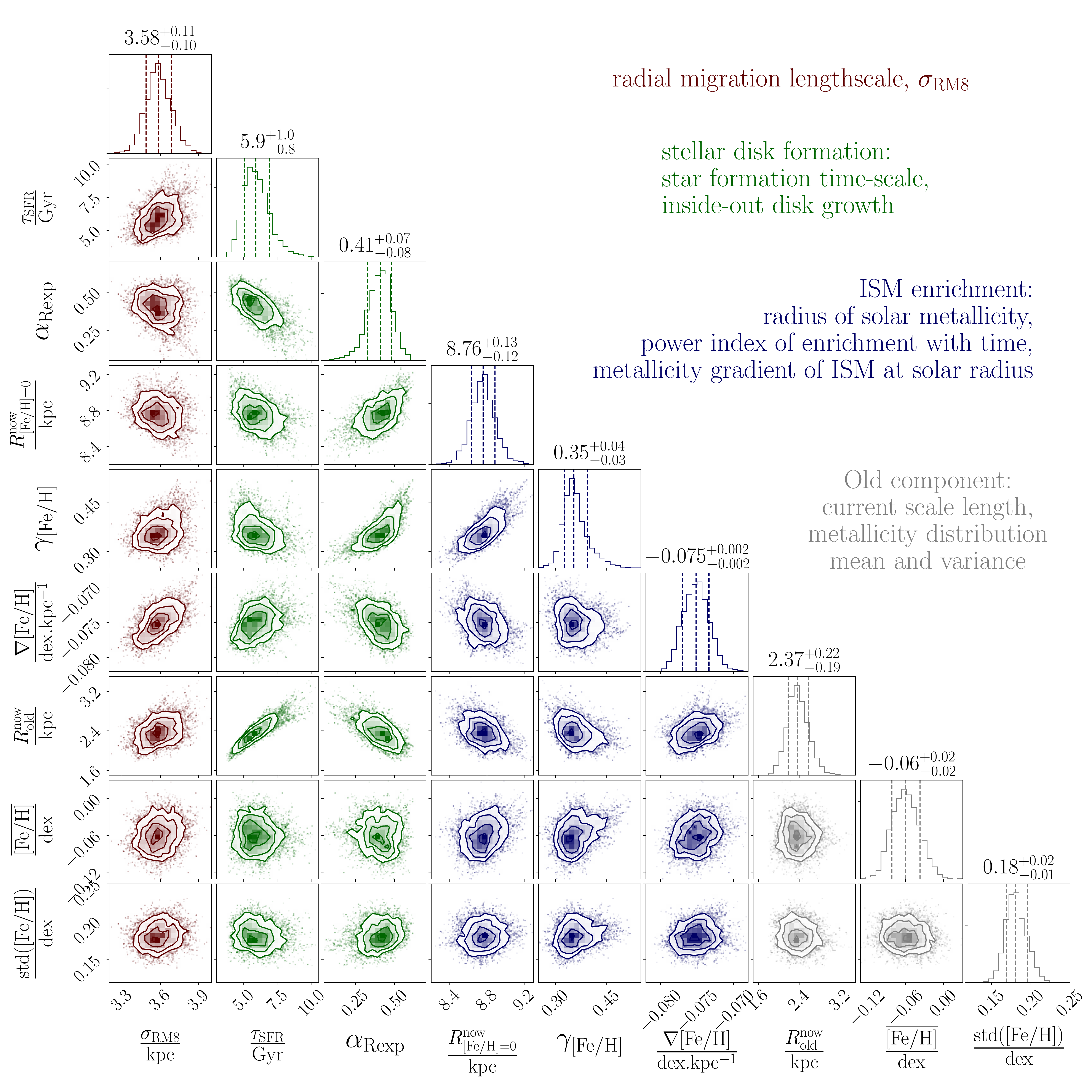}
\caption{\label{fig_corner_full} Posterior distribution of the 9D parameter space. From left to right: the parameter of interest \srm ~for \rom~in kpc (Eq \ref{eq:eq_radial_migration}), followed by the nuisance parameters: star formation time-scale \tsfr ~in Gyr (Eq \ref{eq_age_dist}), the parameter characterizing inside-out disk growth \arexp ~(the Miky Way disk was approximately 40\% smaller at its formation 8 Gyr ago, Eq \ref{eq:eq_radial_birth_profile}). Then come the three parameters characterizing the enrichment of the ISM as a function of time and galactic radius (Eq \ref{eq:eq_Fe_Ro_t}): the radius where the ISM metallicity is solar \rfeh ~in kpc, the power index characterizing the gradual chemical enrichment of the ISM with time $\gfe$, the metallicity gradient of the ISM at the solar radius $\dfe$ in dex $\mathrm{kpc^{-1}}$. The three last parameters are those characterizing the old star: the scale length of the old disk $R_o$ in kpc, mean metallicity of old stars \mfe and its standard deviation \sfe in dex.}
\end{figure*}

\appendix
We include here the full results of the MCMC procedure with all nuisance parameters, including those describing the old low-$\alpha$ stars in the Galactic disk, see Figure \ref{fig_corner_full}. This figure essentially shows that all nuisance parameters are rather well constrained, but there is a degeneracy between the old disk scale-length and the star formation time-scale, as explained in Section \ref{section_results}.

%
%
%
%
%
%

\bibliography{lit}

\end{document}